\documentclass[preprint,authoryear,12pt]{article}
 \usepackage{graphicx}
 \usepackage{epsfig}

\usepackage{bm}
\usepackage{upgreek}

\usepackage{amsmath}
\usepackage{amssymb}
\usepackage{natbib}
\usepackage{color}
\usepackage{tikz}
\usepackage{wrapfig} 
\usepackage{lscape} 
\usepackage{rotating} 

\usepackage[pdftex,bookmarks=true]{hyperref}
\usepackage{authblk}
\graphicspath{{figures/}}

\begin{document}

\title{Evaluating consistency of deterministic streamline tractography in non-linearly warped DTI data}

\author[1]{Nagesh Adluru}
\author[1]{Daniel J. Destiche}
\author[3]{Do P. M. Tromp}
\author[1,4]{Richard J. Davidson}
\author[5]{Hui Zhang}
\author[1,6]{Andrew L. Alexander}

\affil[1]{Waisman Laboratory for Brain Imaging and Behavior, University of Wisconsin-Madison, USA}
\affil[3]{Health Emotions Research Institute, University of Wisconsin-Madison, USA}
\affil[4]{Departments of Psychology \& Psychiatry, University of Wisconsin-Madison, USA}
\affil[5]{Center for Medical Image Computing, Department of Computer Science, University College London, UK}
\affil[6]{Departments of Medical Physics \& Psychiatry, University of Wisconsin-Madison, USA}
\date{}
\maketitle
\begin{abstract}
Tractography is typically performed for each subject using the diffusion tensor imaging (DTI) data in its native subject space rather than in some space common to the entire study cohort. Despite performing tractography on a population average in a normalized space, the latter is considered less favorably at the \emph{individual} subject level because it requires spatial transformations of DTI data that involve non-linear warping and reorientation of the tensors. Although the commonly used reorientation strategies such as finite strain and preservation of principle direction are expected to result in adequate accuracy for voxel based analyses of DTI measures such as fractional anisotropy (FA), mean diffusivity (MD), the reorientations are not always exact except in the case of rigid transformations. Small imperfections in reorientation at individual voxel level accumulate and could potentially affect the tractography results adversely. This study aims to evaluate and compare deterministic white matter fiber tracking in non-linearly warped DTI against that in native DTI. The data present promising evidence that tractography in non-linear warped DTI data could indeed be a viable and valid option for various statistical analysis of DTI data in a spatially normalized space.
\end{abstract}



\section{Introduction}\label{sec:intro}
Diffusion tensor imaging (DTI) \citep{basser_et_al_1994} is the most widely used neuroimaging method for characterizing differences in the microstructural features of brain tissues. However, the anatomic heterogeneity of DTI measures, such as the fractional anisotropy (FA), mean diffusivity (MD) requires strategies for obtaining measurements with high spatial co-localization across individuals \citep{alexander_et_al_2012}. These strategies include manual region-of-interest labeling, tractography-based labeling of specific white matter (WM) pathways, and voxel- or region- based statistical analysis methods that utilize non-linear warping into a standard template space. Each approach has strengths and weaknesses \citep{alexander_et_al_2012}.  

White matter tractography is a powerful algorithmic framework that may be used to estimate and reconstruct the trajectories of WM pathways in the brain using DTI data \citep{mori_et_al_mrm_2002,wakana_et_al_2004,catani_et_al_2008}. Basic deterministic tractography algorithms operate by small step-wise integration of the path following the major eigenvector of the diffusion tensor \citep{mori_et_al_1999,conturo_et_al_1999,alexander_chapter_2011}. Variations of deterministic tractography include higher-order path integration \citep{basser_et_al_2000} and tensor deflection (TEND) \citep{lazar.hbm.2003} which uses the full diffusion tensor to determine the path direction. Such a class of tractography algorithms is highly sensitive to the noise and artifacts in the DTI images, which will lead to errors in tract precision and accuracy \citep{lazar.ni.2003} as well as more catastrophic false branching (false positives) and early termination (false negatives) of tracts. The selection of the stopping criteria has a significant impact on the reconstructed pathways. In particular, the issue of crossing white matter tracts is problematic for tractography reconstruction using DTI.  Despite these limitations, tractography in DTI is widely used to visualize specific white matter pathways, defining specific tracts for region-of-interest analyses \citep{bastin_et_al_2013,eluvathingal_et_al_2007,aoki_et_al_2005}. A main challenge in using tractography for extracting specific pathways is the need for expert labeled way-point filter regions through which the tracts pass and also exclusion filter regions for removing erroneous tracts. Tract selection is usually standardized by first identifying such filter regions on a template and then inverse warping these to the native DTI \citep{thottakara_et_al_2006,lebel_et_al_2008,zhang_et_al_2008} and manual editing. However if one can rely on tractography in the warped DTI such inverse warping and manual editing could be minimized. More recent applications relying on individual level normalized space tractography are in assessing more global structural brain connectivity patterns \citep{hagmann_et_al_2007,chung_et_al_2011,Adluru_et_al_psivt_2012}.

Another common technique used in the analyses of DTI data is spatial normalization, which attempts to non-linearly warp individual DTI data to a DTI template to facilitate strong anatomic and spatial correspondence. Spatial normalization hence forms a key pre-processing step in voxel-based analysis methods \citep{jones_et_al_2002,snook_et_al_2007}. It is also a key step for creating study-specific templates and atlases and also applying standardized templates for region-of-interest analyses \citep{oishi_et_al_2009}. There are many algorithms and strategies for spatial normalization with the most sophisticated being nonlinear diffeomorphic warping with full tensor matching and reorientation. Limitation of spatial normalization is imperfect registration between the DTI data of subjects and the template, and the increased spatial smoothing from the non-linear warping and interpolation.

Spatial normalization has been used to map tractography results into a standardized template space, usually by performing fiber tracking in the un-warped native DTI and then warping the tractography results from native space into the template space \citep{ciccarelli_et_al_2003,johansen_berg_et_al_2005,huang_et_al_2005,hua_et_al_2008,thottakara_et_al_2006}. An alternative strategy is to directly perform tractography in the template space after non-linearly warping the tensors.  Potential advantages of this strategy include performing tractography in a very standard way at the most consistent spatial scale of brain structures. This may be desirable for using tractography to obtain tract-specific regional measurements of DTI measures like FA and MD in normalized space, comparing global brain tractography networks \citep{chung_et_al_2011,Adluru_et_al_psivt_2012} or studying various shape characteristics of the pathways for applications such as classification \citep{adluru_et_al_embs_2009}.

To perform accurate tractography after spatial transformation the diffusion tensors must be reoriented to be consistent with the local anatomic structures. For example, rotations in a rigid body transformation must likewise be applied to the diffusion tensors.  For affine and diffeomorphic (which may be approximated as a piecewise affine) transformations, several strategies have been proposed to reorient the diffusion tensor \citep{alexander_et_al_2001,park_et_al_2003,xu_et_al_2003,ingalhalikar_et_al_2009}. The preservation of principle direction (PPD) method \citep{alexander_et_al_2001} is one of the most widely used strategies. PPD first reorients the major eigenvector direction by the rotation and shear of the local affine transformation followed by similar reorientation of the medium eigenvector. In theory, an accurate transformation with tensor reorientation should maintain the directional and anatomical consistency that is necessary for tractography reconstruction in the normalized space.

Several studies have used spatial normalization with non-linear warping of the tensors with tensor reorientation to create population-averaged diffusion tensor templates, which were used to reconstruct major white matter pathways with tractography \citep{jones_et_al_2002,park_et_al_2003,adluru_et_al_nimg_2012}. Tract reconstructions in the population-averaged DTI data appear much smoother and emphasize the larger, and less complex pathways.  However, relatively few studies have performed tractography on the warped DTI data of the individual subjects \citep{park_et_al_2003,chung_et_al_2011,Adluru_et_al_psivt_2012}. Due to the intricacies involved in warping DTI for analytical inference, we believe it is important to have data to resolve whether tractography performed in the warped tensors is consistent with tractography in the unwarped native tensors in practical situations. An interesting study by \citet{park_et_al_2003} utilized full-white matter tractography in the unwarped and warped DTI to evaluate spatial normalization using different channels (b=0, FA, eigenvalues, full tensor) and found that the full diffusion tensor objective provided the most consistent tractography results between the two spaces. This assumes that tractography in the warped DTI data is not affected by imperfections of the non-linear warping procedure. A key distinction of our study is that we evaluate whether deterministic tractography is consistent despite the imperfections in the normalization process.

Hence, the objective of this study is to evaluate and compare tractography results obtained in native tensors versus those obtained in non-linearly warped tensors. The two key evaluations performed in this study are: (1) Consistency of quantitative DTI (e.g. FA, MD) and volumetric measures of tracts obtained using native and warped tensors. (2) Consistency of geometric shape and anatomical consistency of tractographic reconstructions of pathways using native and warped tensors. A preliminary version of this study appeared in a conference proceeding \citep{adluru_et_al_spie_2013}.

The remainder of the article is organized as follows. We first describe the image acquisition, pre-processing and tract-reconstruction protocols used in this study in Section \ref{sec:materials-methods}. The various evaluations performed are described in Section \ref{sec:comp-eval}. The results of these evaluations are presented and discussed in Section \ref{sec:results} followed by conclusions in Section \ref{sec:conclusion}.

\section{Materials and methods}\label{sec:materials-methods}
\textbf{MR acquisition and basic post-processing:} Diffusion weighted images (DWI) for a set of four healthy adult subjects were acquired using a 3.0-Tesla GE Discovery MR750 MRI scanner (GE Healthcare, Waukesha, WI) and the product 8-channel array head coil with a diffusion-weighted, spin-echo, echo planar imaging sequence.  DW measurements were obtained using 48 non-collinear diffusion encoding directions with diffusion weighting factor of $b = 1000$ s/mm$^2$ in addition to 8 $b = 0$ images. Images were obtained with 2 mm isotropic spatial resolution (field-of-view = 256 mm and 128x128 acquisition matrix) and were interpolated to 1 mm$\times$1 mm in-plane using zero-filled interpolation on the scanner. This does not introduce any blurring but provides visually sharper images on the screen and is the default setting on the GE scanner that is used in many studies. Other parameters were TR = 8 s; TE = 66.2 ms; and echo spacing = 684 $\upmu$s. The DTI protocol was repeated four times, but not averaged on the scanner. Field maps were also acquired using a 2D spoiled gradient echo (SPGR) sequence at two echo times (TE1 = 7 ms; TE2 = 10 ms) to estimate the magnetic field strength across the brain and correct for geometric distortions in the DTI data \citep{jones_cercignani_2010}.  Other parameters were TR = 700 ms, flip angle $\alpha = 60^\circ$, matrix = $256 \times 128$, slice thickness/gap = $4/0$ mm). All the data were corrected for the eddy current related distortion and head motion using $\mathtt{eddy\_correct}$ of FSL software package \citep{jenkinson.2002}. As recommended by \citet{leemans_jones_2009}, the diffusion weighting directions were appropriately rotated using the rotation components of the affine transformations produced as part of the output from the $\mathtt{eddy\_correct}$ command. The distortions from field in-homogeneity were corrected using corresponding field maps. The brain tissue was extracted using the brain extraction tool (BET \citep{smith_bet_2002}), also part of the FSL software package \citep{smith_et_al_fsl_2004}.

\begin{figure}[!htb]
 \centerline{\epsfxsize=1.0\linewidth\epsfbox{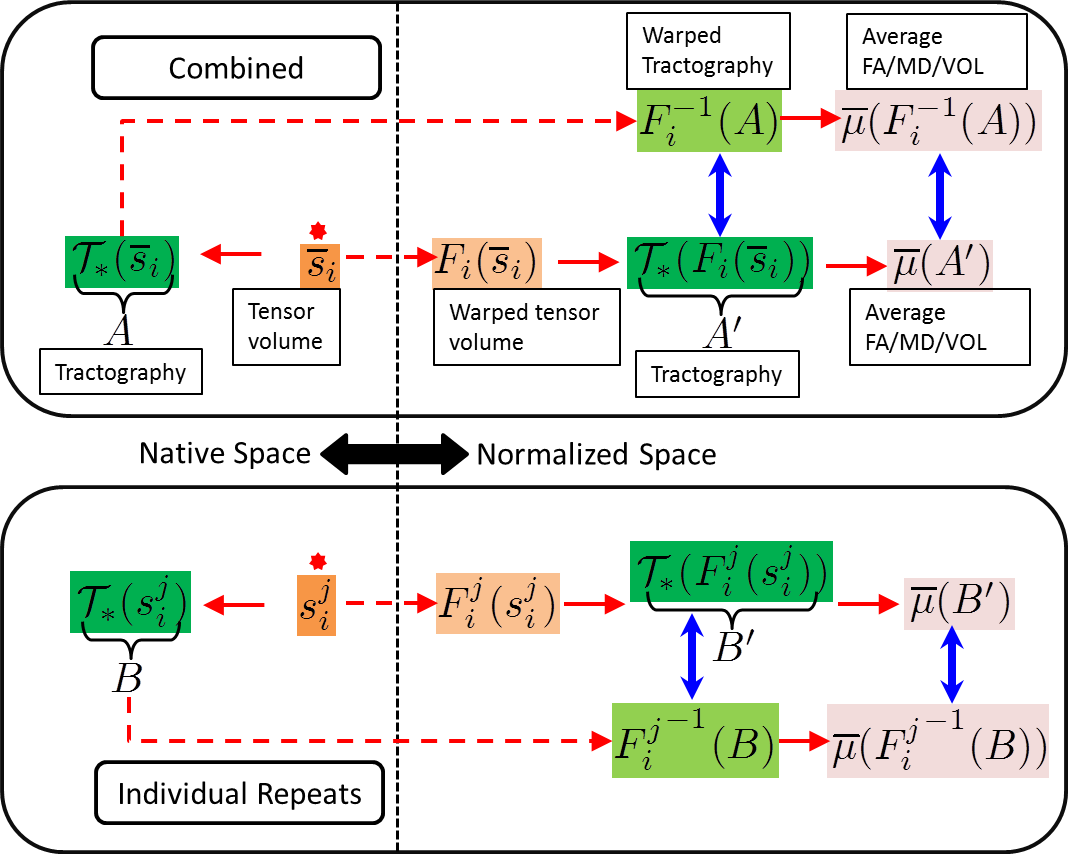}}
\caption{This is a key schematic illustrating the processing done for evaluating the tractography in warped DTI. The red stars indicate the starting points in the data processing for combined and individual repeat data.}\label{fig:comparisons-processing}
\end{figure}

\textbf{DTI processing:} As discussed in the introduction, our goal is to compare white matter fiber tracking in native and warped tensors. Hence we generate the following data per subject as shown in Fig. \ref{fig:comparisons-processing}.
The \textbf{top} panel reflects the data generated by combining the repeated acquisition while the \textbf{bottom} reflects the same for individual repeats. The \textbf{left} of the dotted black line shows the data in native space and the \textbf{right} side shows the corresponding data in the normalized space. The dotted red arrows represent the application of spatial transformations to tensor volumes as well as the tractography data. The color-coding used is as follows: (1) Dark-orange indicates tensor volumes in native space and light orange indicates warped tensor volumes. (2) Dark-green indicates tractography performed on tensor volumes. Light-green reflects the warped tractography data. (3) Light-magenta reflects the quantitative summaries of properties such as FA, MD or VOL obtained. The quantitative summaries ($\overline{\mu}(\mathcal{T})$) are obtained by rasterizing the tractography data into binary masks and then averaging the corresponding measures of voxels in those masks. Rasterization essentially means setting the value of a voxel to one if any part of a tract passes through it. The blue arrows indicate the comparisons made between tractography data as well as their average properties. The symbols are also summarized in the Table \ref{tab:symbols}.
\begin{center}
\begin{table*}[!htb]
\begin{tabular}{|c|p{0.675\linewidth}|}
\hline
\textbf{Symbol} & \textbf{Description}\\
\hline
$\overline{s}_i,s_i^j$&Tensor volume for subject $i$ estimated using combined DWI and $j^{th}$ repeat respectively.\\
$F_i,F_i^j$&Smooth invertible diffeomorphic transformations estimated for warping $\overline{s}_i$ and $s_i^j$ to $\mathcal{A}$ respectively.\\
$A\equiv\mathcal{T}_\ast(\overline{s}_i),B\equiv\mathcal{T}_\ast(s_i^j)$&Fiber reconstructions for subject $i$ by tracking in native DTI volumes.\\
$A'\equiv\mathcal{T}_\ast(F_i(\overline{s}_i)),B'\equiv\mathcal{T}_\ast(F_i^j(s_i^j))$ &Fiber reconstructions for subject $i$ by tracking in warped DTI volumes.\\
$F_i^{-1}(A),{F_i^j}^{-1}(B)$ &Warped $A,B$ into the normalized space respectively.\\
$\overline{\mu}(\mathcal{T})$&Quantitative summaries (FA/MD) and volumetric measures from regions obtained by binarizing the rasterized tractography $\mathcal{T}$.\\
\hline
 \end{tabular}
\caption{Summary of different symbols presented in Fig. \ref{fig:comparisons-processing}. We would like to note that $F$ acts on tensors in the native space while on pointsets in the normalized space. For example, to warp $\overline{s}_i$ which is a tensor in the native space to the normalized space, we apply $F_i$. On the other hand, applying $F_i$ to tracts (which are pointsets) warps them from normalized space to native space. That is $F_i(\overline{s}_i)$ denotes non-linearly warped tensor volume, while $F_i^{\mathbf{-1}}(\mathcal{T}_{\mathtt{\ast}}(\overline{s}_i))$ denotes the non-linearly warped tractography.}\label{tab:symbols}
\end{table*}
\end{center}

\begin{figure}[!htb]
  \centerline{\epsfxsize=0.43\linewidth\epsfbox{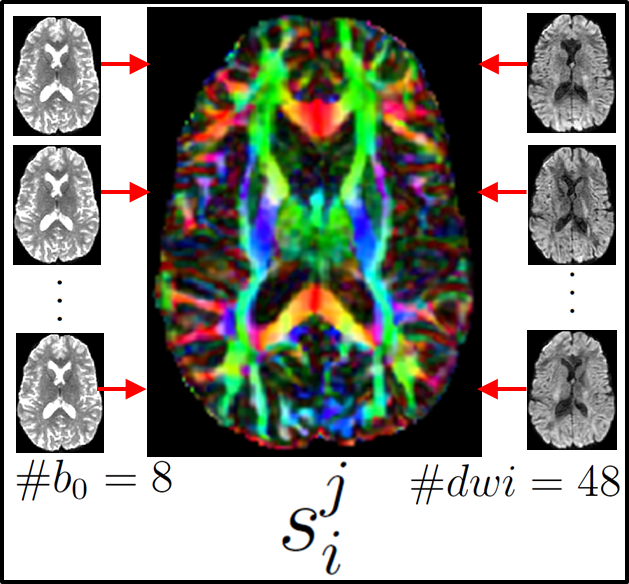}
 \epsfxsize=0.57\linewidth\epsfbox{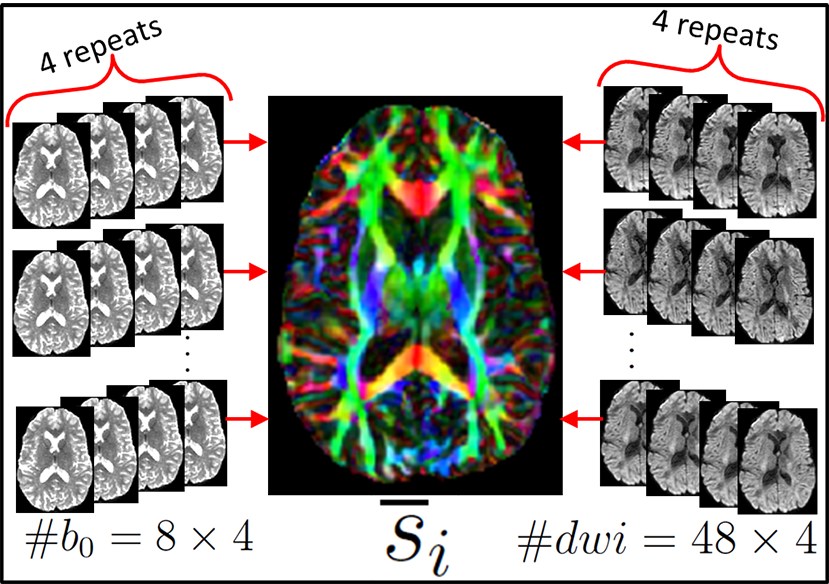}}
\caption{This figure illustrates the two different types of tensor estimation performed in the study. \textbf{Left:} For each subject 8 $b$0s and 48 DWI volumes are used in tensor volume estimation for each of the four individual repeats. The motion correction step is performed for each repeat independently by registering all the images to the corresponding volume's first $b$0 image. \textbf{Right:} The four repeats (32 $b$0s and 192 DWI volumes) are combined by averaging resulting in a higher-quality tensor estimation. Here the motion correction is performed by registering all the volumes of all the repeats to the first $b$0 of the first repeat.}\label{fig:individual_combined}
\end{figure}

The tensor volumes were estimated using non-linear estimation using Camino \citep{camino_cook:2006}, for all the 16 individual repeat scans as well as by combining four repeats as shown in Fig. \ref{fig:individual_combined}.

\begin{figure}[!htb]
  \centerline{\epsfxsize=0.5\linewidth\epsfbox{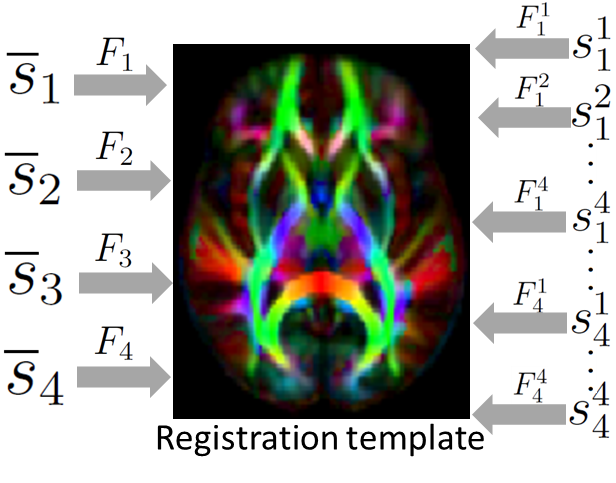}}
\caption{This figures summarizes the spatial transformations estimated using DTI-TK. The 16 different tensor volumes (four individual repeats for four different subjects) and the 4 combined tensor volumes are registered to a template obtained from a different study with the same acquisition protocol. The set of transformations $F$ are stored and used in the creation of the study specific atlas ($\mathcal{A}$) as shown in Eq. \eqref{eq:atlas}.}\label{fig:registration_atlas}
\end{figure}

\textbf{DTI warping:} The tensor volumes estimated from the individual repeats as well as from the combined scans were non-linearly warped to a common DTI template which was obtained from scans (with the same acquisition parameters) of a different set of 16 healthy subjects from a similar age group. This mimics a standard setting of registering a population to a standardized template. This procedure is summarized in Fig. \ref{fig:registration_atlas}. The template was estimated iteratively using rigid, affine and diffeomorphic warping using DTI-TK \citep{zhang.2006} which has been shown to be an effective registration method for DTI data \citep{wang_et_al_2011,adluru_et_al_nimg_2012}. Finally the population average or atlas ($\mathcal{A}$) for this study is obtained by log-Euclidean averaging \citep{Arsigny_et_al_2005} the non-linearly warped individual repeat data as,
\begin{equation}
 \mathcal{A}=\textrm{exp}\left(\frac{\overset{4}{\underset{i=1}{\sum}}\overset{4}{\underset{j=1}{\sum}}\log(F_i^j(s_i^j))}{16}\right),\label{eq:atlas}
\end{equation}
where $\textrm{exp}$ and $\log$ are the standard matrix exponentiation and matrix logarithm operators. We would like to note that although $\{F_i^j\}_{j=1}^4$ and $F_i$ are expected to be identical there could be minor differences based on the amount of motion between the repeat scans.

\textbf{Tractography protocol:} For each individual repeat (16 in total) and each combined scan (4 in total), we reconstructed four major WM pathways using tractography performed with both the original ($s$) and non-linearly warped tensor volumes ($F(s)$). The tractography was performed on the whole brain using FACT algorithm implemented in Camino \citep{camino_cook:2006}. The seed file for tractography was the white matter mask defined as FA$>$0.2 and the stopping criteria were either FA$<$0.2 or a maximum curvature threshold of $60^\circ$. The step size for the tracking algorithm is 1/10$^{\textrm{\textrm{th}}}$ the slice thickness which is 2 mm. The structures examined were corpus callosum ($\mathtt{CC}$), bilateral cingulum bundle ($\mathtt{CB}$), bilateral inferior fronto-occipital fasciculus ($\mathtt{IFO}$) and bilateral uncinate fasciculus ($\mathtt{UF}$). These structures were extracted by systematically following protocols described in \citet{wakana_et_al_2004,catani_et_al_2008}. Since the way-point and exclusion $\mathtt{ROI}$s are difficult to delineate accurately using FA maps alone, tractography data was used to iteratively refine the $\mathtt{ROI}$s. Tracts were converted to Trackvis software \citep{wang_et_al_2007} format for visualization purposes using a conversion tool\footnote{\url{http://www.nitrc.org/projects/camino-trackvis/}}. Each structure was extracted by applying the $\mathtt{ROI}$s to whole brain tractography, denoted symbolically as
\begin{equation}
\mathcal{T}_\mathtt{struct}(s)\equiv \mathbb{F}\left(\mathcal{T}_\mathtt{WB}(s),\mathtt{ROI_{struct}}(s)\right), 
\end{equation}
where $\mathbb{F}$ denotes filtering of tracts using a set of way-point and exclusion $\mathtt{ROI}$s needed for the specific structure and $\mathcal{T}_{\mathtt{WB}}(s)$ denotes whole brain ($\mathtt{WB}$) tractography obtained by tracking in the tensor volume $s$ using FACT tracking algorithm and EULER interpolation. For all the $\mathtt{struct}$s, rather than delineating the required way-point and exclusion $\mathtt{ROI}$s in the individual subject, they are identified on the atlas ($\mathcal{A}$) and either inverse warped into the native space or applied directly in the normalized space. For example, the uncinate fasciculus ($\mathtt{UF}$) tracts in the tensor volumes $\overline{s}_i$ and $F_i(\overline{s}_i)$ are obtained respectively as
\begin{align}
\mathcal{T}_\mathtt{UF}(\overline{s}_i)&\equiv\mathbb{F}\left(\mathcal{T}_\mathtt{WB}(\overline{s}_i),F_i^{-1}(\mathtt{ROI_{UF}}(\mathcal{A}))\right),\\
\mathcal{T}_\mathtt{UF}(F_i(\overline{s}_i))&\equiv\mathbb{F}\left(\mathcal{T}_\mathtt{WB}(F_i(\overline{s}_i)),\mathtt{ROI_{UF}}(\mathcal{A})\right).
\end{align}
This minimizes the amount of manually introduced inconsistencies of the $\mathtt{ROI}$s and variability in the comparisons.

\begin{figure}[!htb]
 \centerline{\epsfxsize=0.5\linewidth\epsfbox{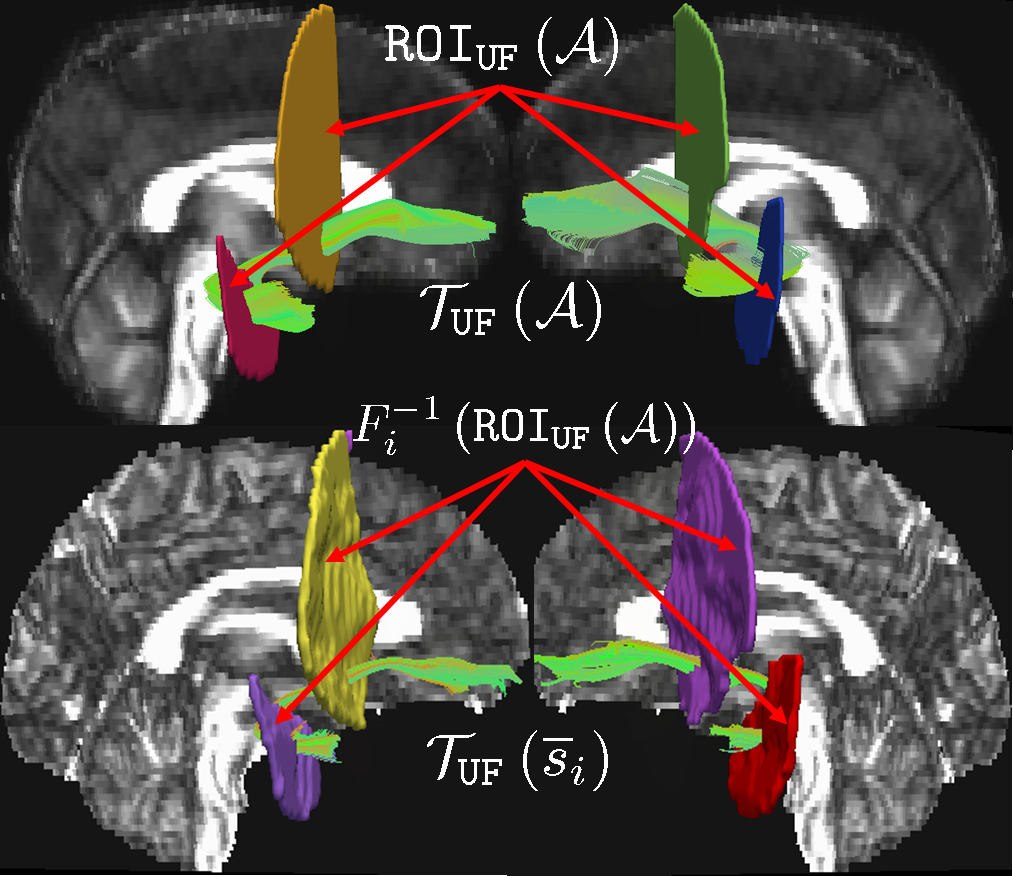}}
\caption{Sample reconstruction of uncinate fasciculus ($\mathtt{UF}$) using waypoint $\mathtt{ROI}$s manually identified on the atlas and inverse warped onto individual subject data. Two way point $\mathtt{ROI}$s are identified on each side of the brain in the frontal region of the brain near amygdala. The underlays are the corresponding FA maps.}\label{fig:UNC_recon}
\end{figure}

As an example, below we describe and demonstrate extraction of the $\mathtt{ROI}$s for $\mathtt{UF}$ briefly. $\mathtt{UF}$ belongs to the family of association fibers (also considered to belong to the limbic system) with significant projects to the fronto-orbital cortical areas from the temporal lobe. These are short range U shaped fibers and are one of the major amygdalo-prefrontal pathways which are implicated in anxiety \citep{tromp_et_al_2012}. Delineation of the $\mathtt{UF}$ required the definition of $\mathtt{ROI}$s by identifying the most posterior coronal slice that showed clear separation of the frontal and temporal lobes bilaterally. As depicted in Fig. \ref{fig:UNC_recon}, $\mathtt{ROI}$s were manually drawn around the frontal lobes and around the temporal lobe on the identified coronal view of FA on the atlas ($\mathcal{A}$). These are identified on both left and right hemispheres of the brain. Then, the Boolean AND term was used to only include fibers that crossed through the frontal and temporal $\mathtt{ROI}$s for each hemisphere separately.
\subsection{Consistency evaluations}\label{sec:comp-eval}
Two sets of quantitative evaluations were performed: (1) geometric shape consistencies of the tracts and (2) consistency of quantitative DTI and volumetric measures of the tracts. The geometric shape consistency comparisons matter most for TSA \citep{yushkevich.2008,zhang_et_al_2010}, network analysis \citep{chung_et_al_2011,Adluru_et_al_psivt_2012} or other feature extraction based analyses such as classification \citep{adluru_et_al_embs_2009} while the consistency of quantitative DTI and volumetric measures are most relevant for region-of-interest analyses.

The consistency of shapes of the tracts was measured using dice similarity coefficients and $\kappa$ scores \citep{cohen_kappa_1960}. Although there are myriad number of ways for capturing shape consistencies, dice similarity coefficients and $\kappa$ scores are widely used in comparing shapes in constrained settings such as brain image analysis where rotational and other transformational invariances are not needed. The dice similarity coefficient which reflects percentage volume overlap between the pairs of masks obtained from the two sets of tractography, $\mathcal{T}_1$, $\mathcal{T}_2$ is computed as
\begin{equation}
 \mathtt{DC}(\mathcal{T}_1,\mathcal{T}_2)=\frac{2M(\mathcal{T}_1)\cap M(\mathcal{T}_2)}{|M(\mathcal{T}_1)|+|M(\mathcal{T}_2)|},
\end{equation}
where $M(\mathcal{T}_1)$ and $M(\mathcal{T}_2)$ are the binary masks obtained by rasterizing the tractography sets $\mathcal{T}_1$ and $\mathcal{T}_2$, respectively. The $\kappa$ score, which is defined based on agreement between expected and observed data \citep{landis_koch_1977} is computed as
\begin{equation}
 \kappa(\mathcal{T}_1,\mathcal{T}_2)=\frac{\mathtt{obs}-\mathtt{exp}}{1-\mathtt{exp}},
\end{equation}
where
\begin{align}
 \mathtt{obs}=\mathtt{nn+pp},&\quad\mathtt{exp}=\mathtt{Enn+Epp},\\
 \mathtt{nn}=|M(\mathcal{T}_1)=0~ \& ~M(\mathcal{T}_2=0)|,&\quad\mathtt{pp}=|M(\mathcal{T}_1)=1~ \& ~M(\mathcal{T}_2=1)|,\\
\mathtt{np}=|M(\mathcal{T}_1)=0~ \& ~M(\mathcal{T}_2=1)|,&\quad\mathtt{pn}=|M(\mathcal{T}_1)=1~ \& ~M(\mathcal{T}_2=0)|,\\
\mathtt{Enn}=\frac{(\mathtt{nn+pn})(\mathtt{nn+np})}{\mathtt{N}},&\quad\mathtt{Epp}=\frac{(\mathtt{pp+pn})(\mathtt{pp+np})}{\mathtt{N}},\\
&\mathtt{N=nn+np+pn+pp}.
\end{align}
Dice coefficient could be considered to imply a good shape overlap when $>$0.7. The $\kappa$ scores on the other hand are typically interpreted in terms agreement according to \citet{landis_koch_1977} criteria as slight [0.11$\leq\kappa<$0.2), fair [0.2$\leq\kappa<$0.4), moderate [0.4$\leq\kappa<$0.6), substantial [0.6$\leq\kappa<$0.8) and almost perfect [0.8$\leq\kappa\leq$1.0).

Since the ground-truth measurements of the quantitative summary measures of the structures are unknown, the consistency of these properties were visualized using Bland-Altman (BA) plots \citep{altman_bland_1983,bland_altman_1986}. BA plots are used to measure agreement between two methods both of which have sources of uncertainty. They visualize the dependence of the bias on the average. The BA plots show scatter plots of the averages of the individual subject on the abscissa and the differences of the individual subject on the ordinate,
\begin{equation}
 \mathtt{BA}_i(x,y)=\left(\frac{m_1(i)+m_2(i)}{2},m_1(i)-m_2(i)\right),
\end{equation}
where $m_1(i)$ is $\overline{\mu}(A')$ or $\overline{\mu}(B')$ for $i^{th}$ subject and $m_2(i)$ is $\overline{\mu}(F_i^{-1}(A))$ or $\overline{\mu}(F_i^{-1}(B))$ respectively (see Fig. \ref{fig:comparisons-processing} and Table \ref{tab:symbols}). If there is no systematic variance in the bias according to the average and if the individual subject biases in the metrics cluster around the total bias, then a high degree of agreement between two different methods could be implied.

Concordance correlation coefficients ($\rho_c$) were used as a summary statistic since BA plots convey qualitative information. The concordance coefficient is an adjusted Pearson correlation ($\rho$) and is defined as \citep{lin_1989},
\begin{equation}
 \rho_c=\frac{2\rho \sigma_{m_1}\sigma_{m_2}}{\sigma_{m_1}^2+\sigma_{m_2}^2+(\mu_{m_1}-\mu_{m_2})^2},\label{eq:concordance}
\end{equation}
where $\mu_{m_1}$ and $\sigma_{m_1}$ are the inter-scan/subject mean and standard deviations of the average microstructural properties obtained using $A$ or $B$. $\mu_{m_2}$ and $\sigma_{m_2}$ reflect those obtained using $F_i^{-1}(A)$ or $F_i^{-1}(B)$.

\section{Results}\label{sec:results}
\subsection{Geometric shape consistency}
\begin{figure}[!htb]
\centerline{\epsfxsize=1.0\linewidth\epsfbox{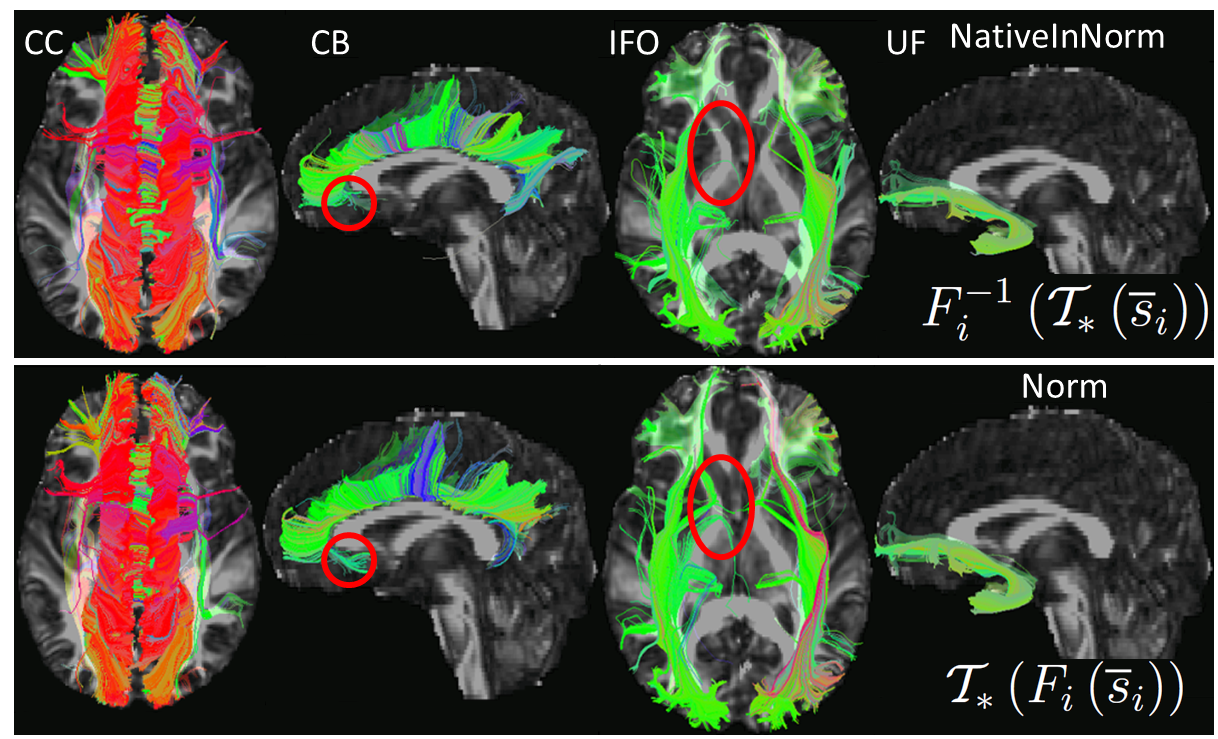}}
\caption{Visual consistency of the tractography based reconstructions by tracking in un-warped native (top) and non-linearly warped tensors (bottom). The red circles highlight some minor inconsistencies between the reconstructions of cingulum bundle ($\mathtt{CB}$) and the inferior fronto-occipital ($\mathtt{IFO}$).}\label{fig:vis_consistency_tracts}
\end{figure}

\begin{figure}[!htb]
\centerline{\epsfxsize=1.0\linewidth\epsfbox{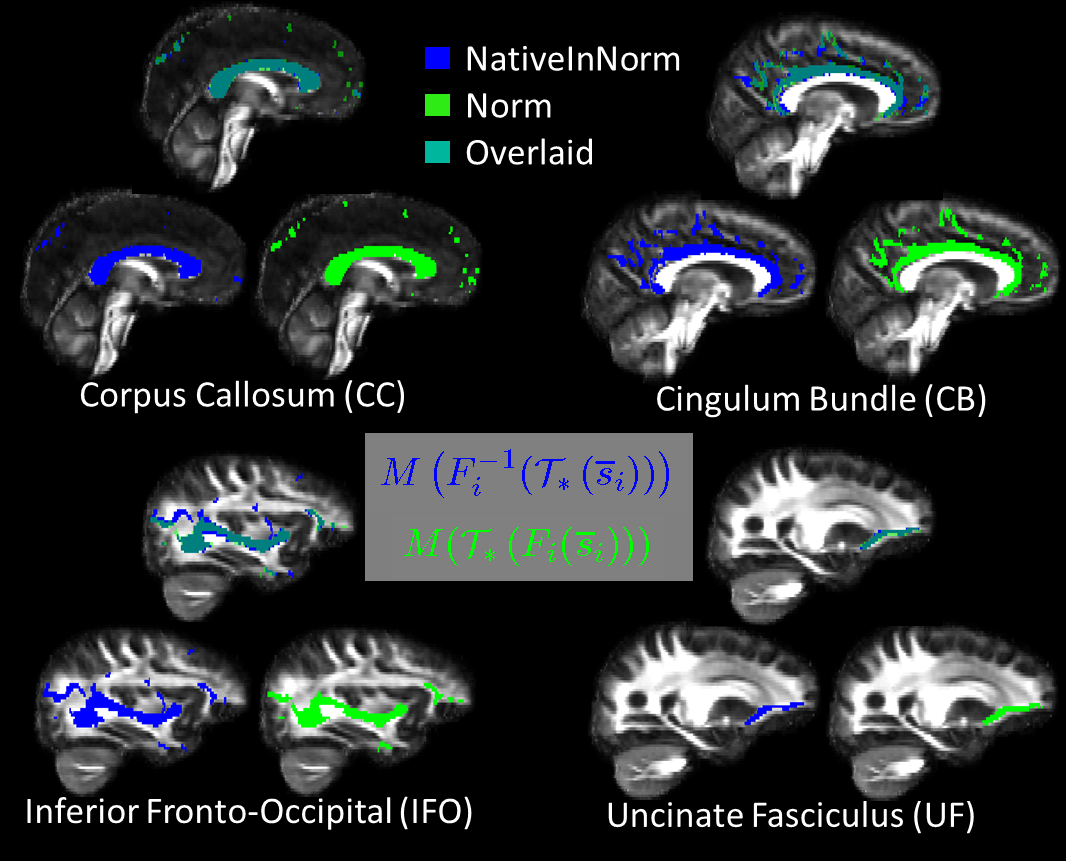}}
\caption{Visual consistency of the rasterized masks ($M(\mathcal{T})$) obtained from the tract reconstructions by tracking in native and non-linearly warped tensors in the normalized space.}\label{fig:vis_consistency_tracts_mask}
\end{figure}

Figs. \ref{fig:vis_consistency_tracts} and \ref{fig:vis_consistency_tracts_mask} present the tract reconstructions and the masks created by rasterizing them, respectively. These figures show high level of visual consistency between the tract reconstructions showing that tracking in non-linearly warped tensors is not substantially affected in reconstructing major white matter structures such as the ones considered here.

\begin{figure}[!htb]
\centerline{\epsfxsize=0.425\linewidth\epsfbox{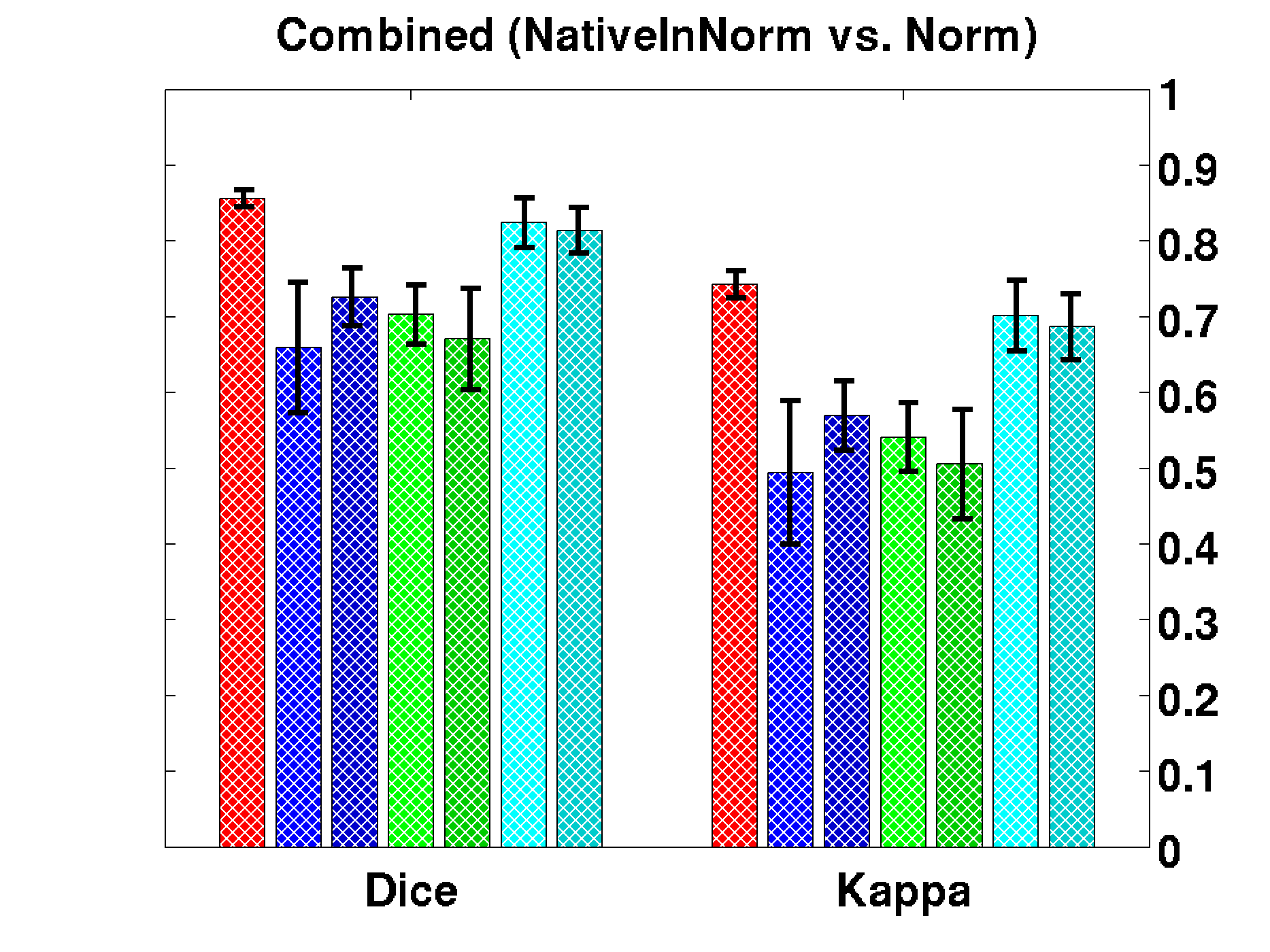}
\epsfxsize=0.425\linewidth\epsfbox{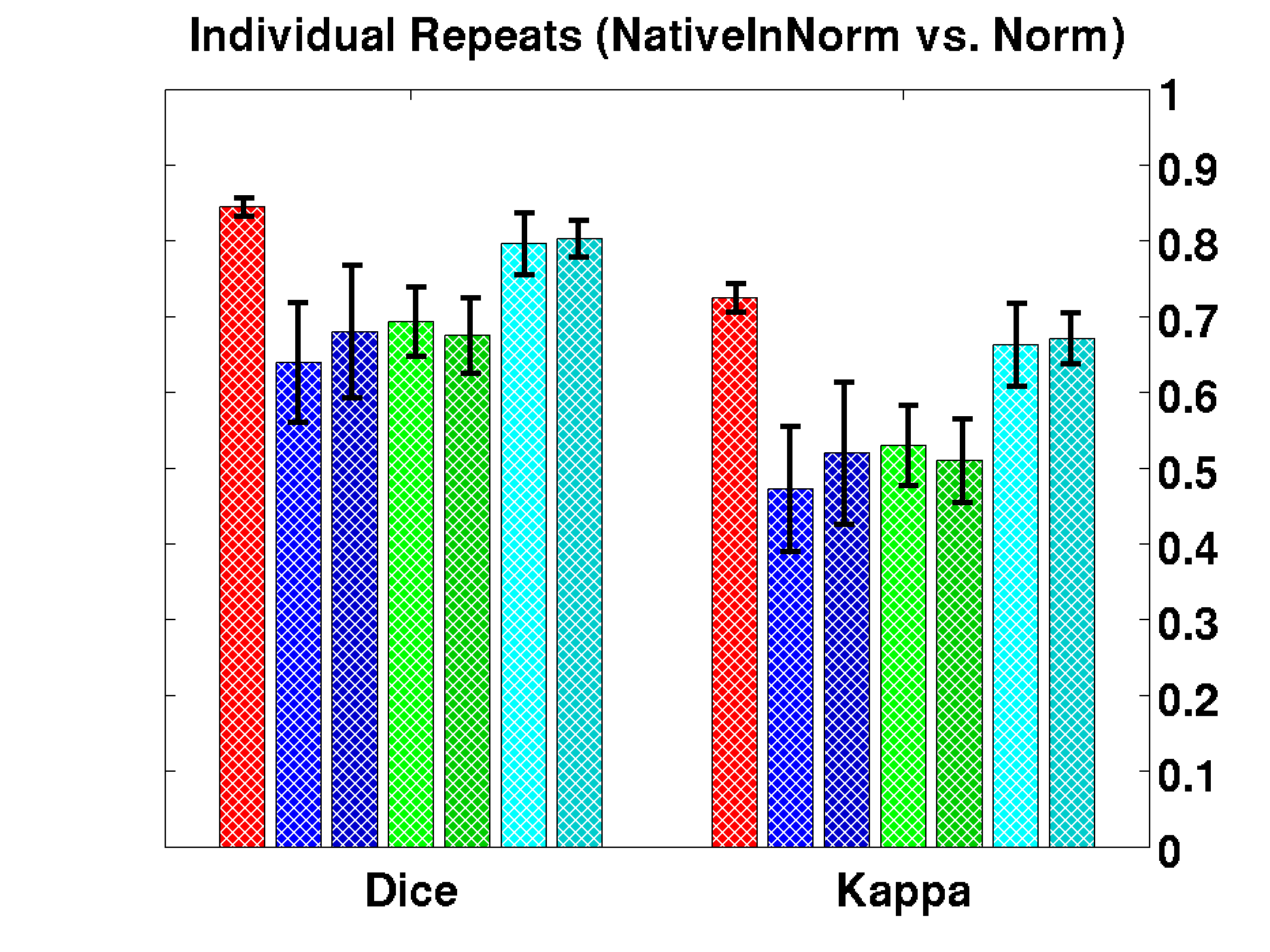}
\epsfxsize=0.15\linewidth\epsfbox{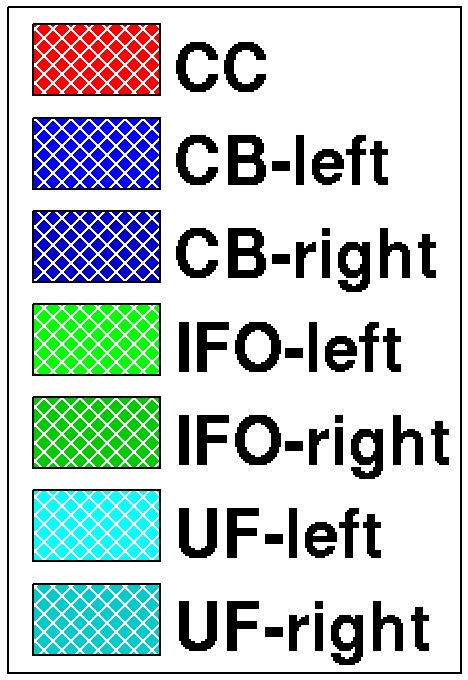}}
\centerline{\makebox[0.425\linewidth][c]{\textbf{(a)}} \makebox[0.425\linewidth][c]{\textbf{(b)}}
\makebox[0.15\linewidth]{}}
\caption{Geometric consistency of the tracts obtained by tracking in native and non-linearly warped tensors estimated using (a) combined as well as (b) individual repeat DWI data. The consistencies are measured via dice coefficient and $\kappa$ scores. We show the consistency of the left and right tracts separately and note that as expected, there is no significant difference between the two sides.}\label{fig:kappa_dice_method_bar_plots}
\end{figure}

Fig. \ref{fig:kappa_dice_method_bar_plots} plots the dice and $\kappa$ scores for shape consistency of different tracts. Although we observe that the scores indicate good shape overlaps and all the structures show consistent patterns in terms of shape agreement, the scores were lower for $\mathtt{CB}$ and $\mathtt{IFO}$. $\mathtt{CB}$ is a long, narrow tract adjacent to the corpus callosum (one of the largest tracts in the brain) and is dominated by branching along the entire path. The worse scores are also likely caused by partial volume averaging with $\mathtt{CC}$ and surrounding gray matter. $\mathtt{IFO}$ similarly is also a long and narrow tract with branching. The lower scores hence could to be attributed to the difficulty in their localization rather than to the effects of tracking in non-linear warped of tensors.

\begin{figure}[!htb]
\centerline{\epsfxsize=0.5\linewidth\epsfbox{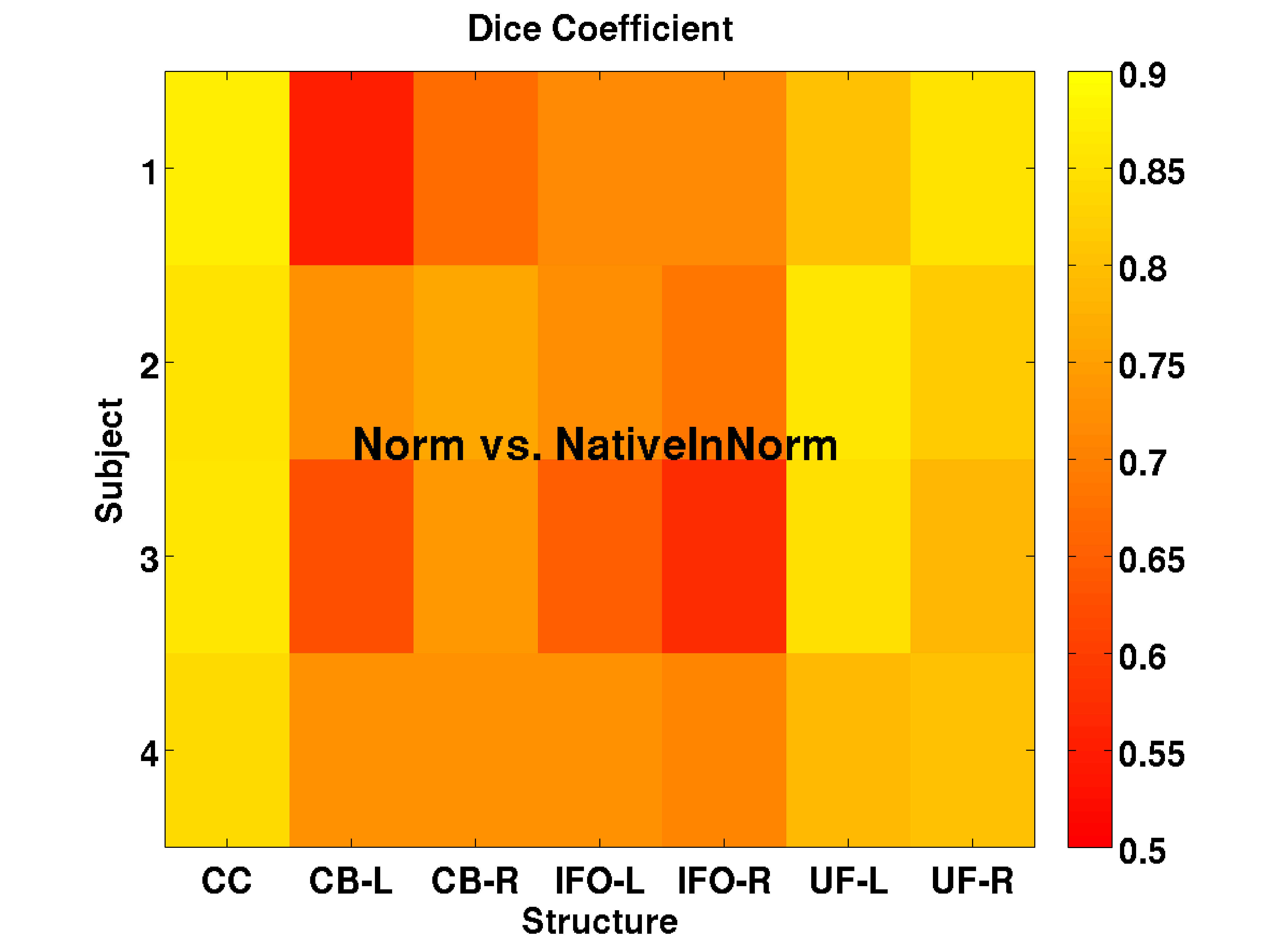}
\epsfxsize=0.5\linewidth\epsfbox{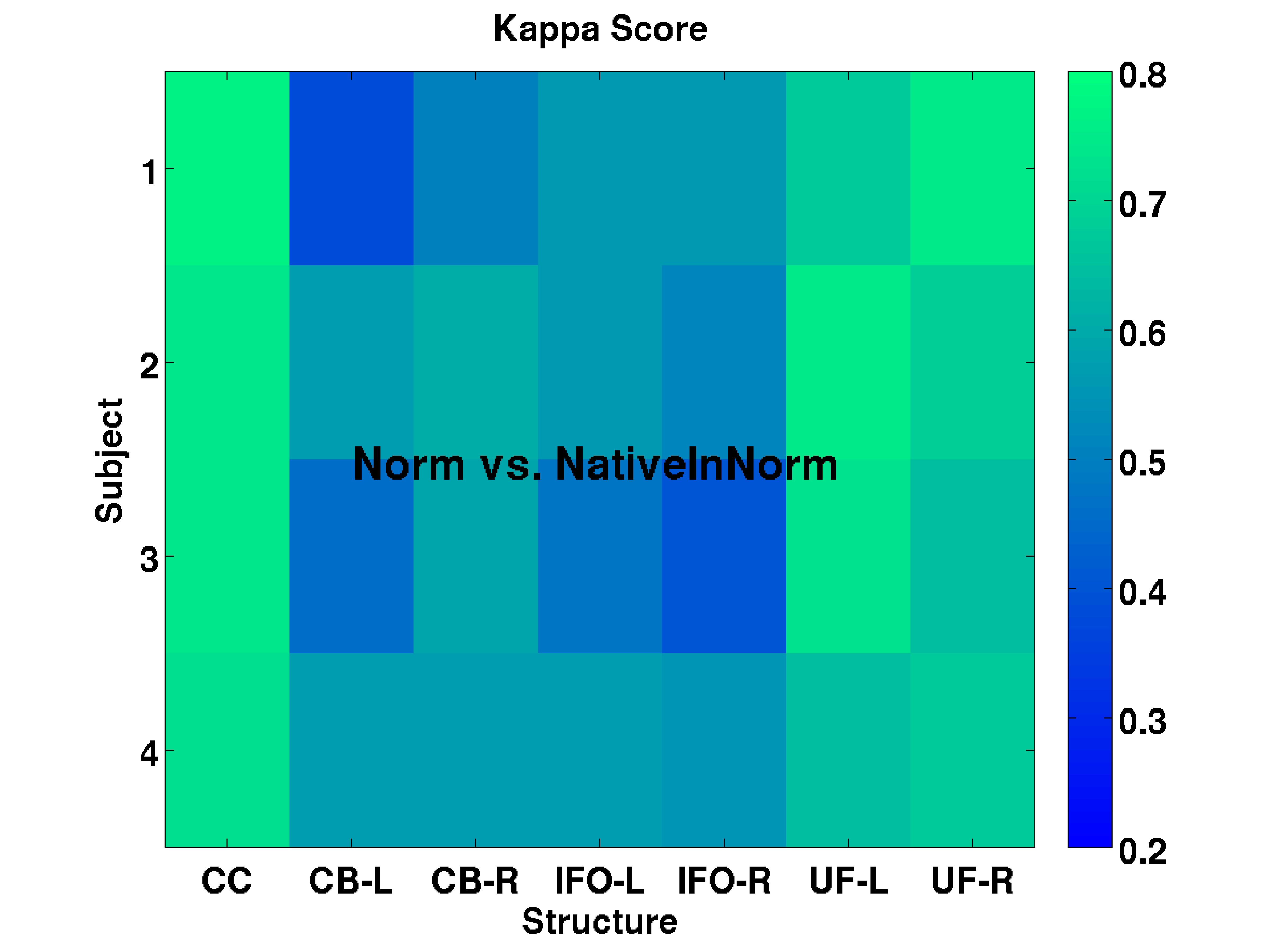}}
\centerline{\epsfxsize=0.5\linewidth\epsfbox{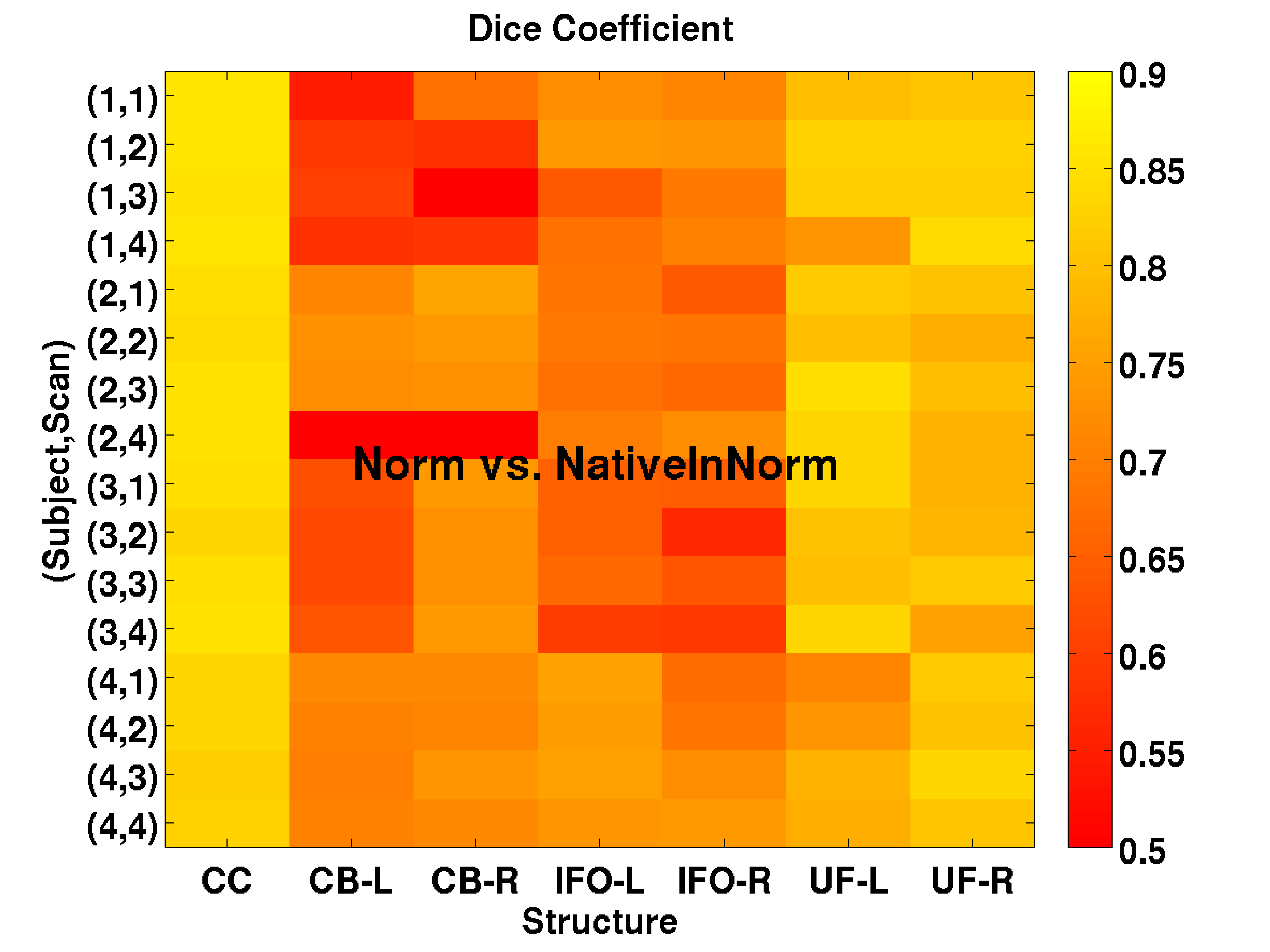}
\epsfxsize=0.5\linewidth\epsfbox{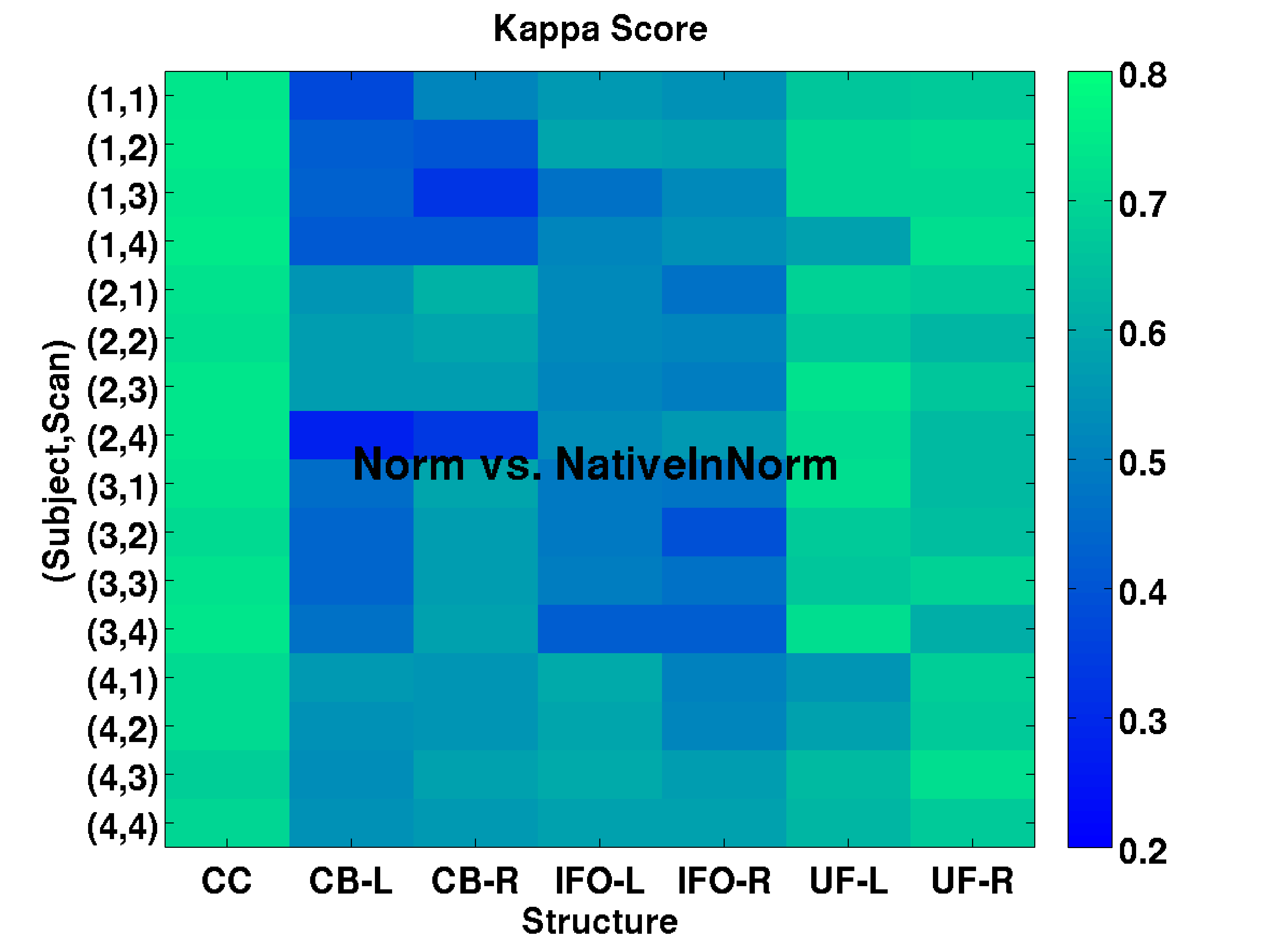}}
\caption{Inter-subject (top) and inter-scan (bottom) variance in the dice coefficients and the $\kappa$ scores between tracts (including left and right for CB, IFO, UF) obtained by traking in non-linearly warped and un-warped native tensors.}\label{fig:kappa_dice_structure_imagesc}
\end{figure}

Fig. \ref{fig:kappa_dice_structure_imagesc} shows the inter-subject and inter-scan variation of the dice coefficients and $\kappa$ scores for all the structures. As discussed above the scores for $\mathtt{CB}$ and the $\mathtt{IFO}$ tracts are lower but as seen from the consistent score distributions in these maps, non-linear warping of tensors does not induce any additional substantial noise variance in the shapes of structures in the individual subjects.

\begin{figure}[!htb]
\centerline{\epsfxsize=\linewidth\epsfbox{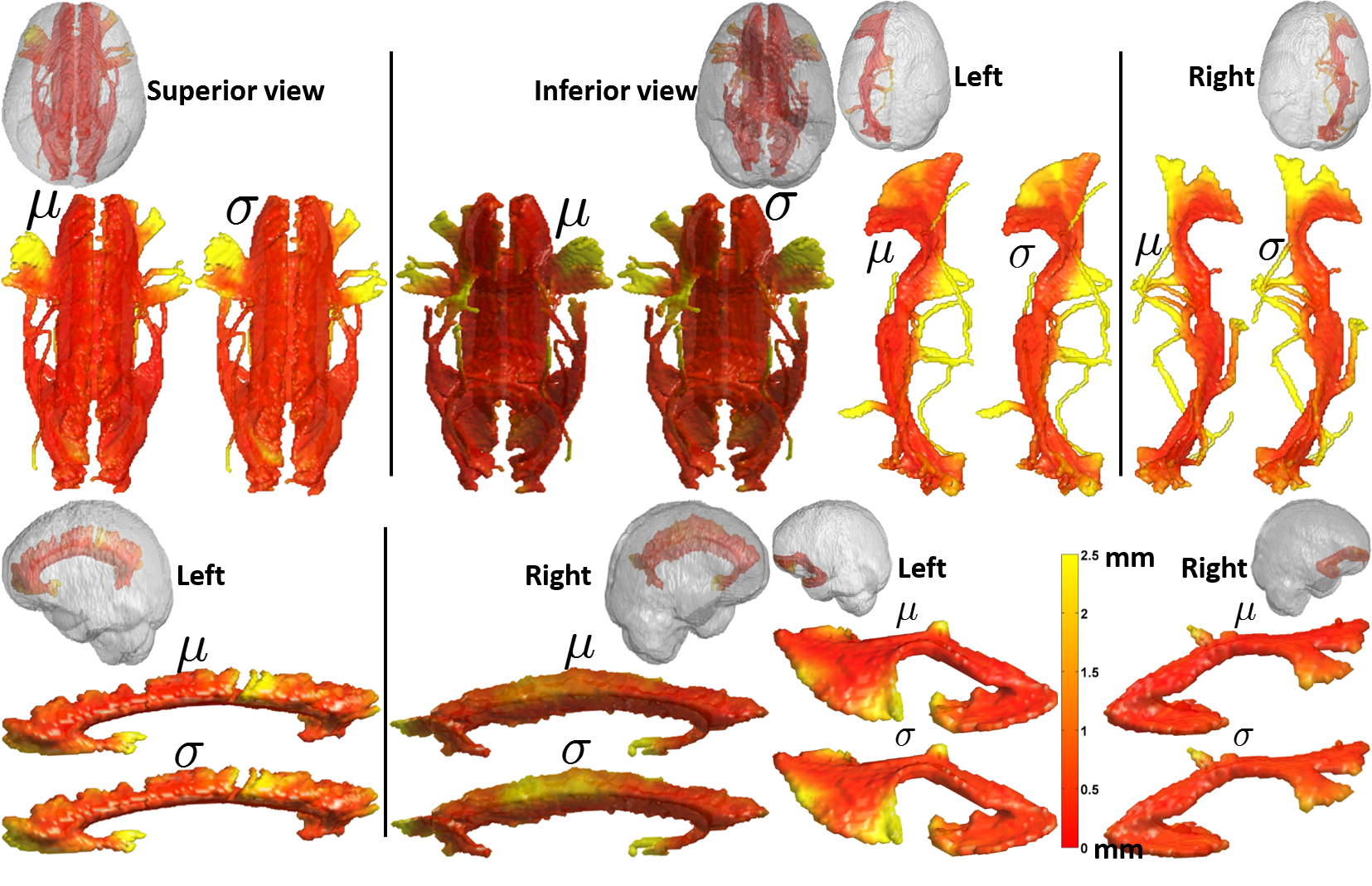}}
\caption{Shape consistency fluctuation maps for the four tracts. CC and IFO are shown on the top row left and right respectively while CB and IFO are shown in the bottom row left and right respectively. We can observe that the flucutuations in the consistency of the tracts between native and normalized space tracking are mostly concentrated on the "outskirts" of the tracts. The fluctuations in the majority of the tract regions are less than 1mm. We would like to recall that the voxel resolution used in our experiments is 2mm isotropic with 1mm$\times$1mm zero-filled in plane resolution which suggests that the shape of the tracts obtained by native and normalized space tracking is quite consistent.}\label{fig:shape_consistency_fluctuation}
\end{figure}

In addition we also present the shape consistency fluctuation maps in the Fig. \ref{fig:shape_consistency_fluctuation} for all the four tracts. As opposed to the dice and $\kappa$ scores these maps reveal the spatial dependence of the overlap scores. For each tract, these maps are obtained as follows. We first fit an isosurface to each of the binary masks ($M(T)$) obtained by native and normalized space tracking. The isosurface is also fit to the binary masks obtained by tracking on the atlas as well. For each vertex ($v_i^\mathcal{A}$) on the atlas tract surface we obtain a distance measure ($d_i$) as,
\begin{equation}
d_i=\arrowvert\underset{j=1\cdots N}{min}~d(v_i^\mathcal{A},v_j^\textrm{Native})-\underset{j=1\cdots N'}{min}~d(v_i^\mathcal{A},v_j^\textrm{Norm})\arrowvert,\label{eq:scfm}
\end{equation}
where $N$ and $N'$ are the number of triangular faces on the surfaces obtained from native and normalized space tracking respectively. The distance between two vertices ($d(v_i,v_j)$) is simply the the Euclidean distances between them. The $d_i$s are computed for each of the 16 comparisons (using the 16 scans) and the mean and standard deviations are mapped to a color scale and the tract surfaces are visualized to show how the consistency fluctuates along different parts of the tract.
\subsection{Consistency of quantitative DTI and volumetric measures}\label{sec:micro-results}
\begin{figure}[!htb]
\centerline{\epsfxsize=0.33\linewidth\epsfbox{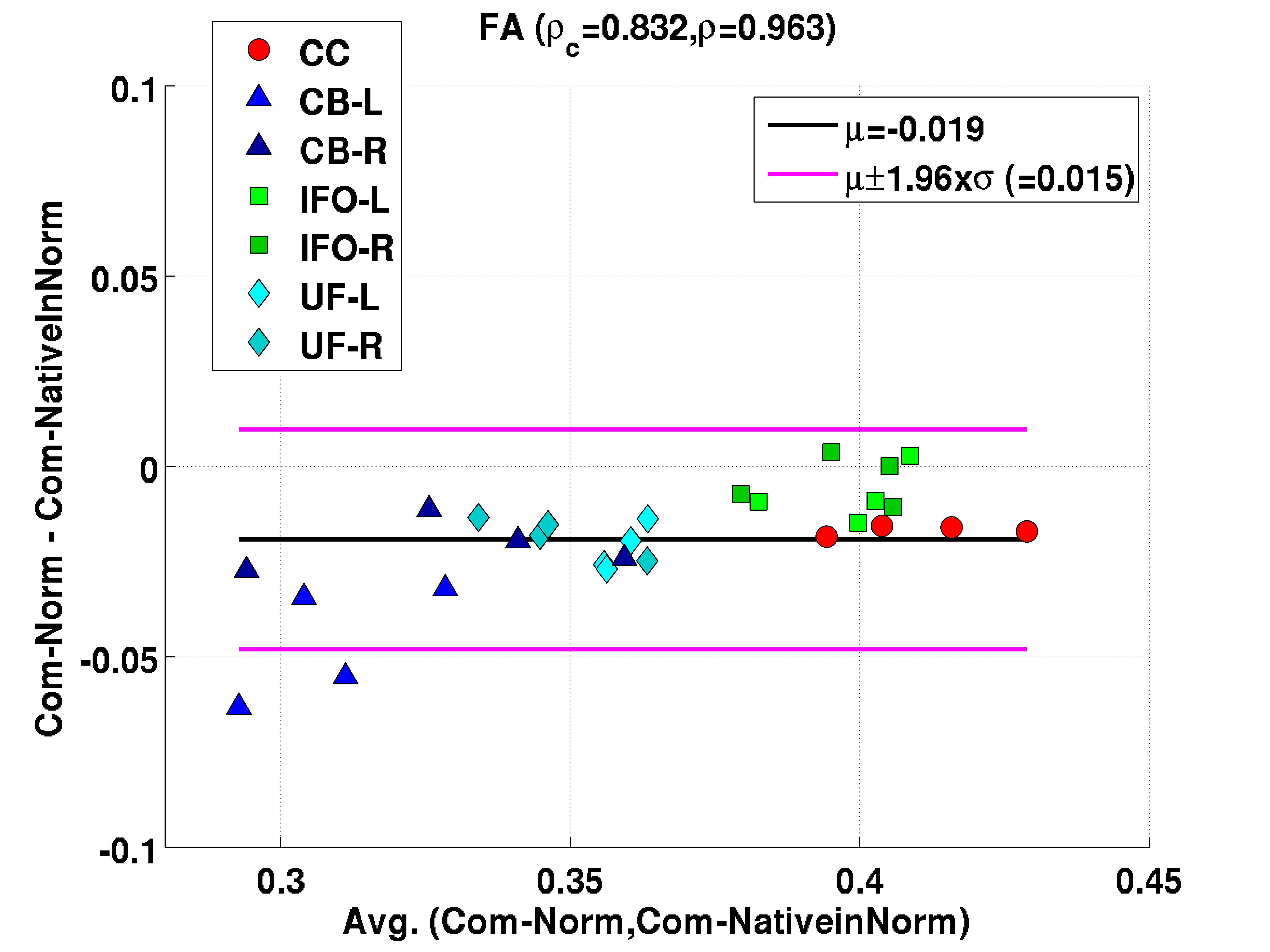}
\epsfxsize=0.33\linewidth\epsfbox{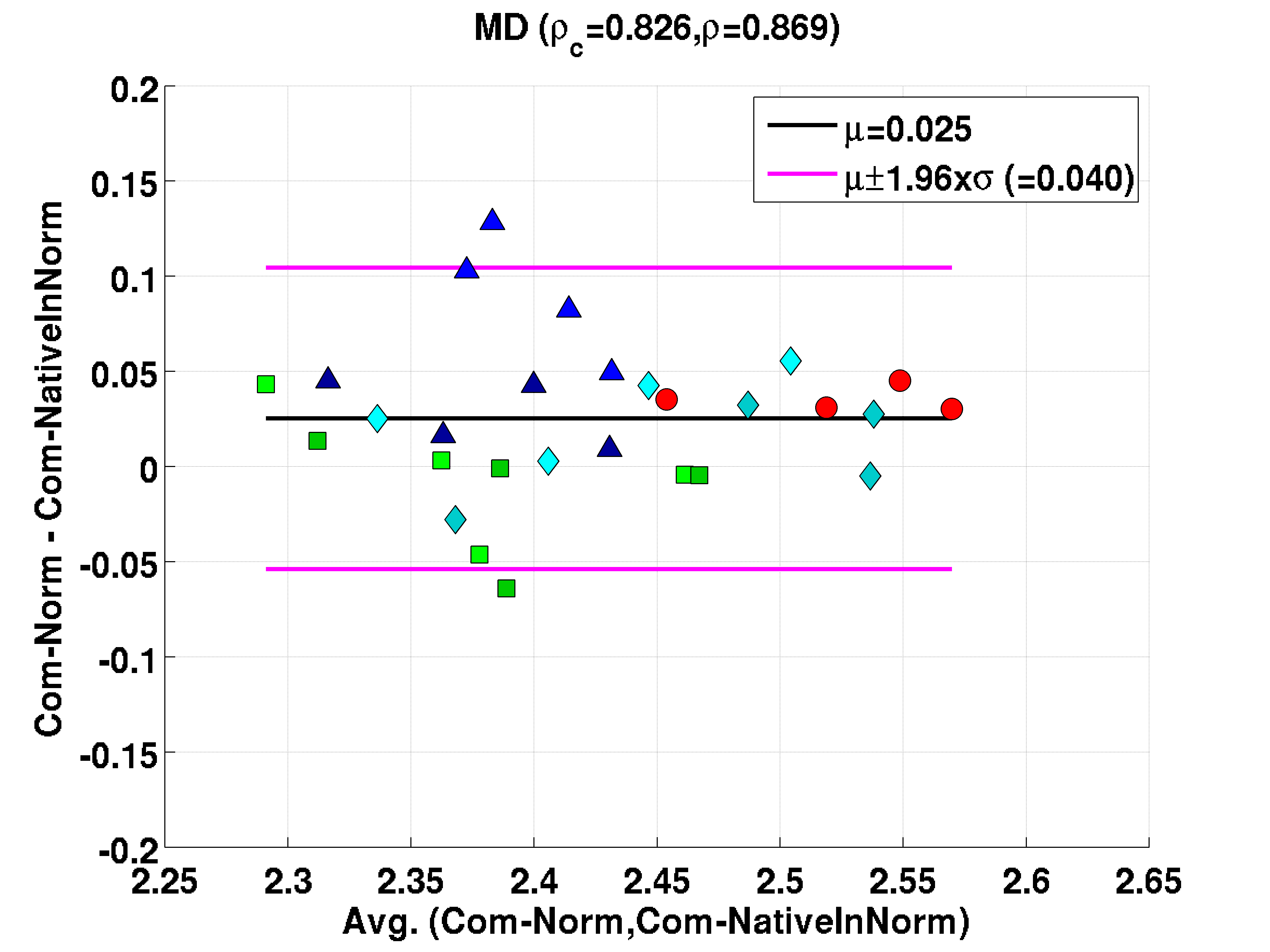}
\epsfxsize=0.33\linewidth\epsfbox{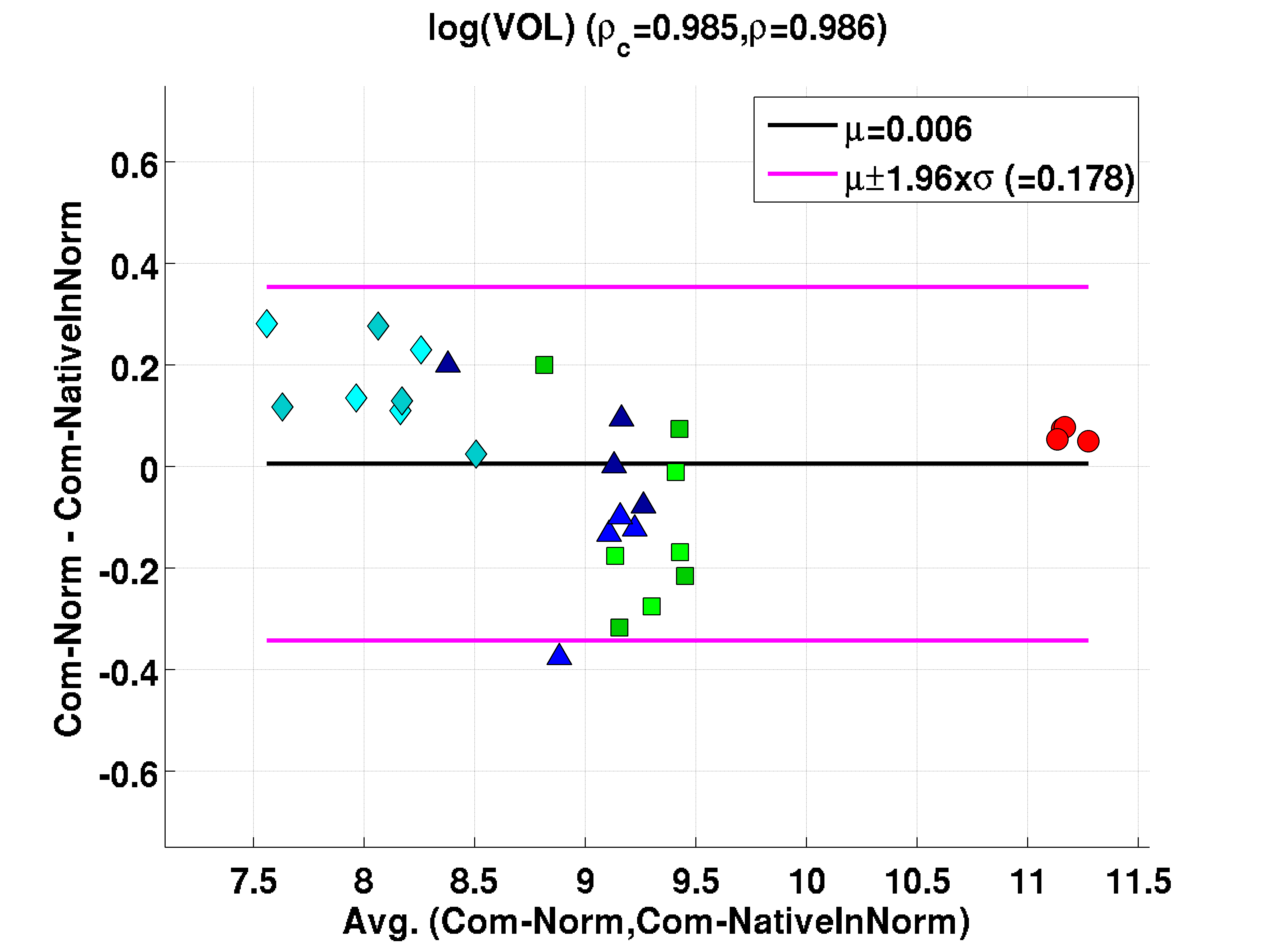}}
\centerline{\epsfxsize=0.33\linewidth\epsfbox{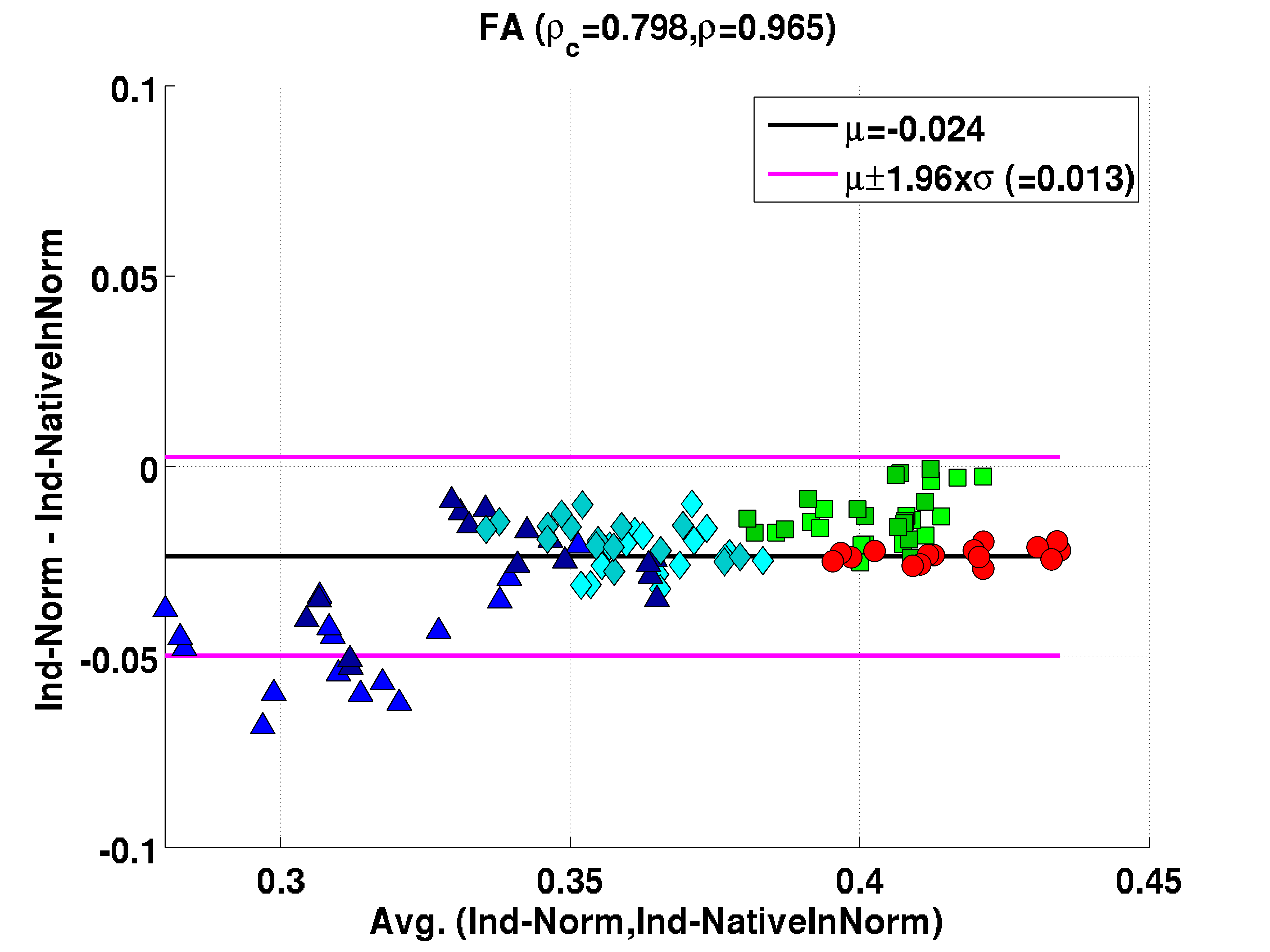}
\epsfxsize=0.33\linewidth\epsfbox{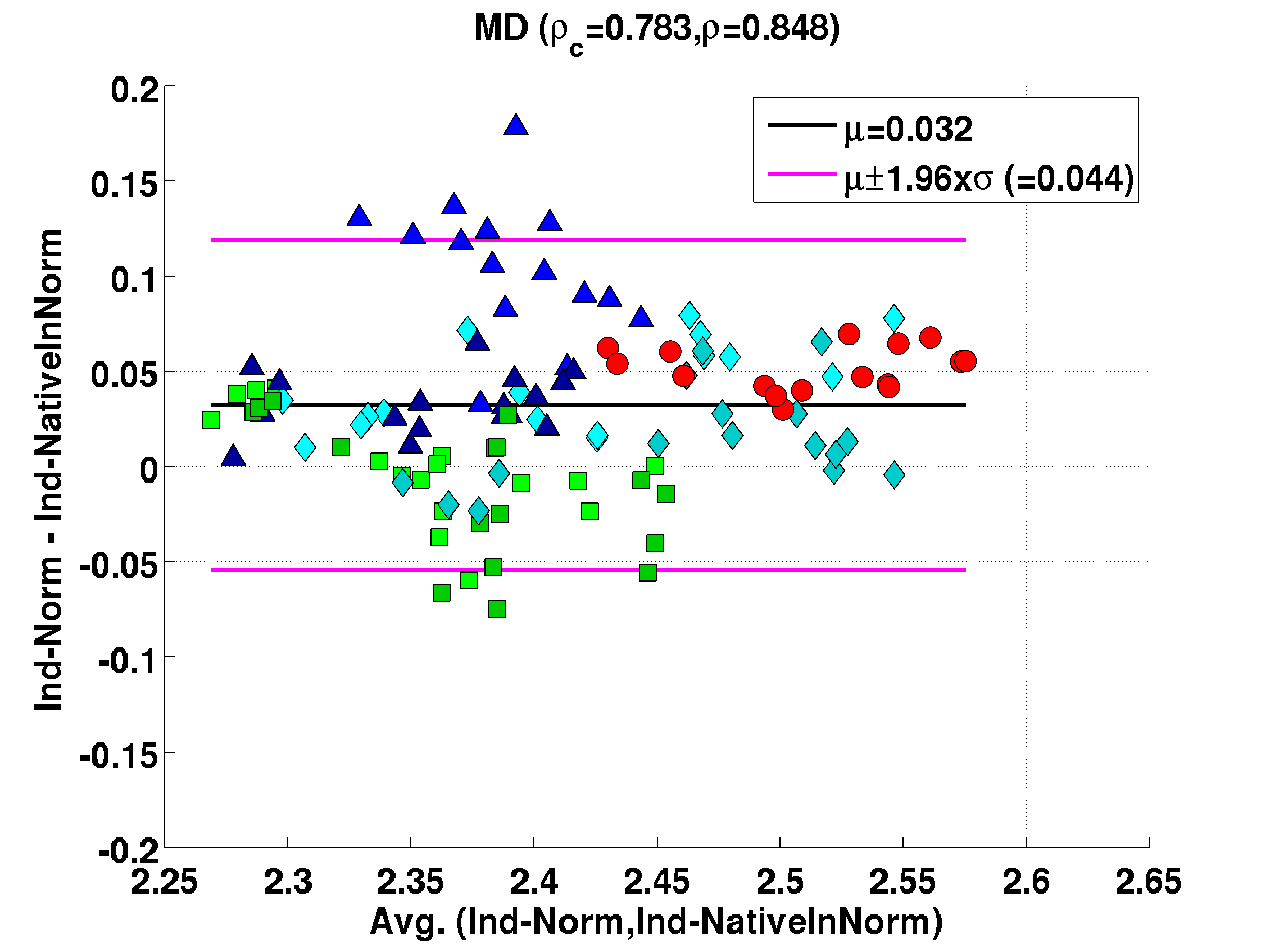}
\epsfxsize=0.33\linewidth\epsfbox{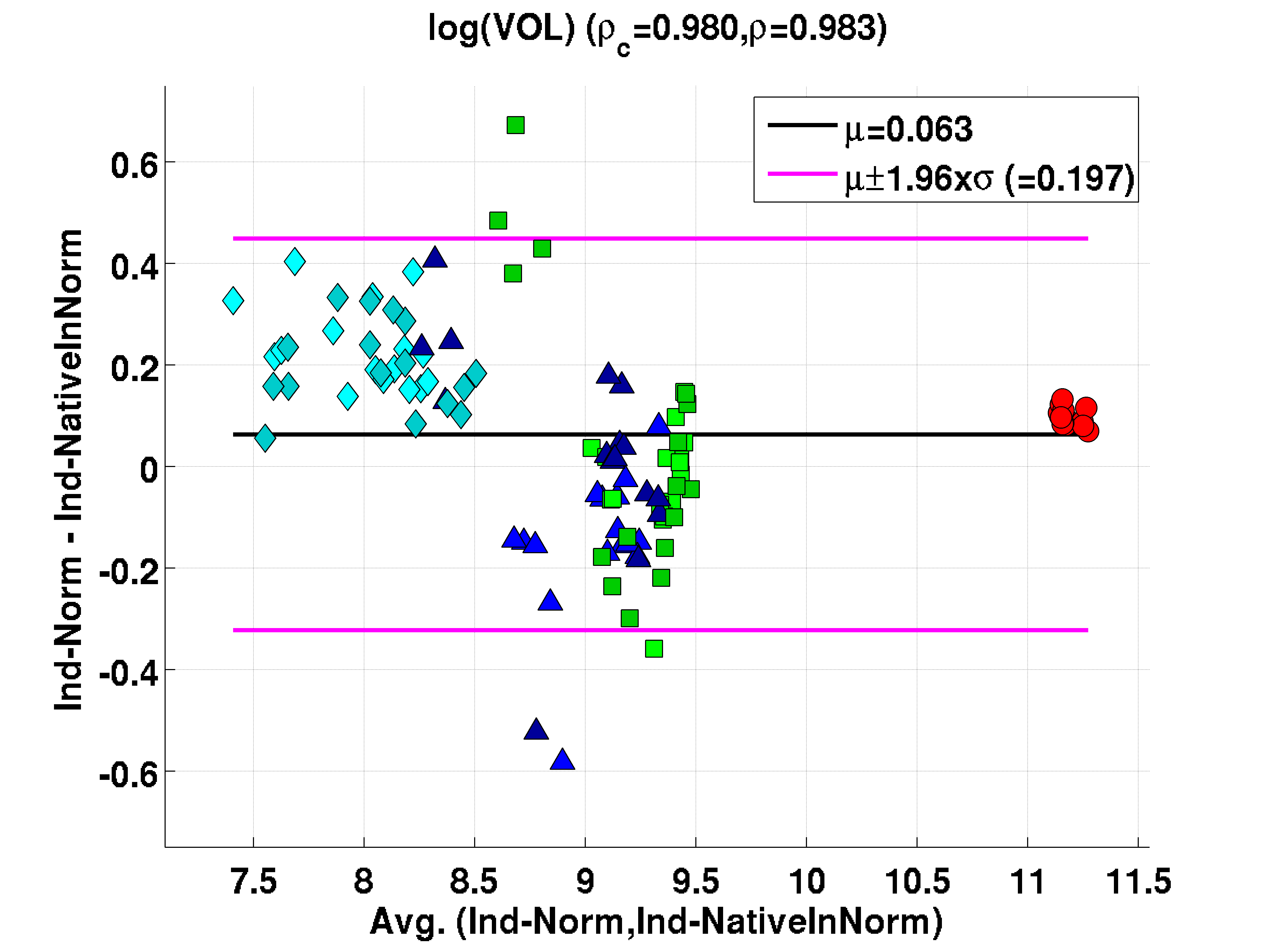}}
\caption{Bland-Altman plots to measure the agreements between the quantitative summary measures of the four tracts (including left and right for the CB, IFO and UF) ($\overline{\mu}(\mathcal{T})$) by tracking in native and non-linearly warped tensors -- fractional anisotropy (FA) (left), mean diffusivity (MD) in $\upmu$m$^2$/s units (center) and $\log$(volume) (VOL) in mm units (right). The means ($\mu$) and the standard deviations ($\sigma$) of the differences, the concordance coefficients ($\rho_c$) and Pearson correlation coefficients ($\rho$) (displayed in the titles) are estimated using the summaries from all the structures jointly. The top row shows the results for tensors estimated from combined DWI while the bottom row for tensors estimated from individual repeat DWI.}\label{fig:fa_tr_vol_ba_basic_consistency_1}
\end{figure}

The BA plots are shown in Fig. \ref{fig:fa_tr_vol_ba_basic_consistency_1}. These plots reveal that the individual subject differences are clustered around the means and mostly fall within $1.96$ standard deviations of the means suggesting that that the distributions are normal and also that there is a high-level of agreement between the measures of structures obtained by tracking in native and warped tensors. We also observe that overall differences in the quantitative and volumetric measures are small and consistent for the different structures.

For FA the overall absolute difference is on the order of 0.019 for combined data and 0.024 for individual repeat data. For MD it is 0.025 $\upmu$m$^2$/s when using combined data and 0.035 $\upmu$m$^2$/s when using individual repeat data. The overall difference in $\log$(VOL) is 0.006 mm and 0.068 mm when using combined and individual repeat data respectively. Since the difference for VOL was estimated on the log-scale it reflects the ratio of the volumes rather than their difference.

The Pearson correlation coefficients are above 0.96 for FA and $\log$(VOL) and on the order of 0.85 for MD. This reflects a high degree of agreement between the variances of the measures obtained using tractography in the native and warped tensors. The concordance coefficients are on the order of 0.8 for FA and MD and 0.98 for $\log$(VOL). The reduction in the concordance coefficients for FA and MD compared to the Pearson correlation is because of the differences in their means ($\mu_{m_1}-\mu_{m_2}$) (Eq. \eqref{eq:concordance}).

Although a few data points of the scatter falls outside the standard deviation interval, we can observe that there is no systematic pattern in the differences, except that the structures with smaller quantitative and volumetric measures tend to have larger variation in the difference when tracking in unwarped vs. warped tensors. Interestingly we also observe that in case of FA the overall variation in the differences is smaller compared to that in cases of MD and $\log$(VOL). This is consistent with the fact that the diffusion tensor warping algorithms such as DTI-TK \citep{zhang.2006} minimize the distance metrics that emphasize the anisotropic components of the tensors. The plots also reveal, as expected, that any sort of averaging (combining individual repeats) or interpolations due to spatial warping is likely to result in a smoothing effect (increasing the region covered by the tracts) albeit only by small amount. Such an effect of warping has also been reported recently in \cite{bockhorst_et_al_2012}.

The overall absolute difference in FA, MD, $\log$(VOL) is consistently higher when using individual repeat data than when using combined data. This is consistent with an empirically known fact that noise in the DWI data inflates the FA of the tensors \citep{Pierpaoli_and_Basser_1996,Jones_and_Basser_2004}.

\begin{figure}[!htb]
\centerline{\epsfxsize=0.33\linewidth\epsfbox{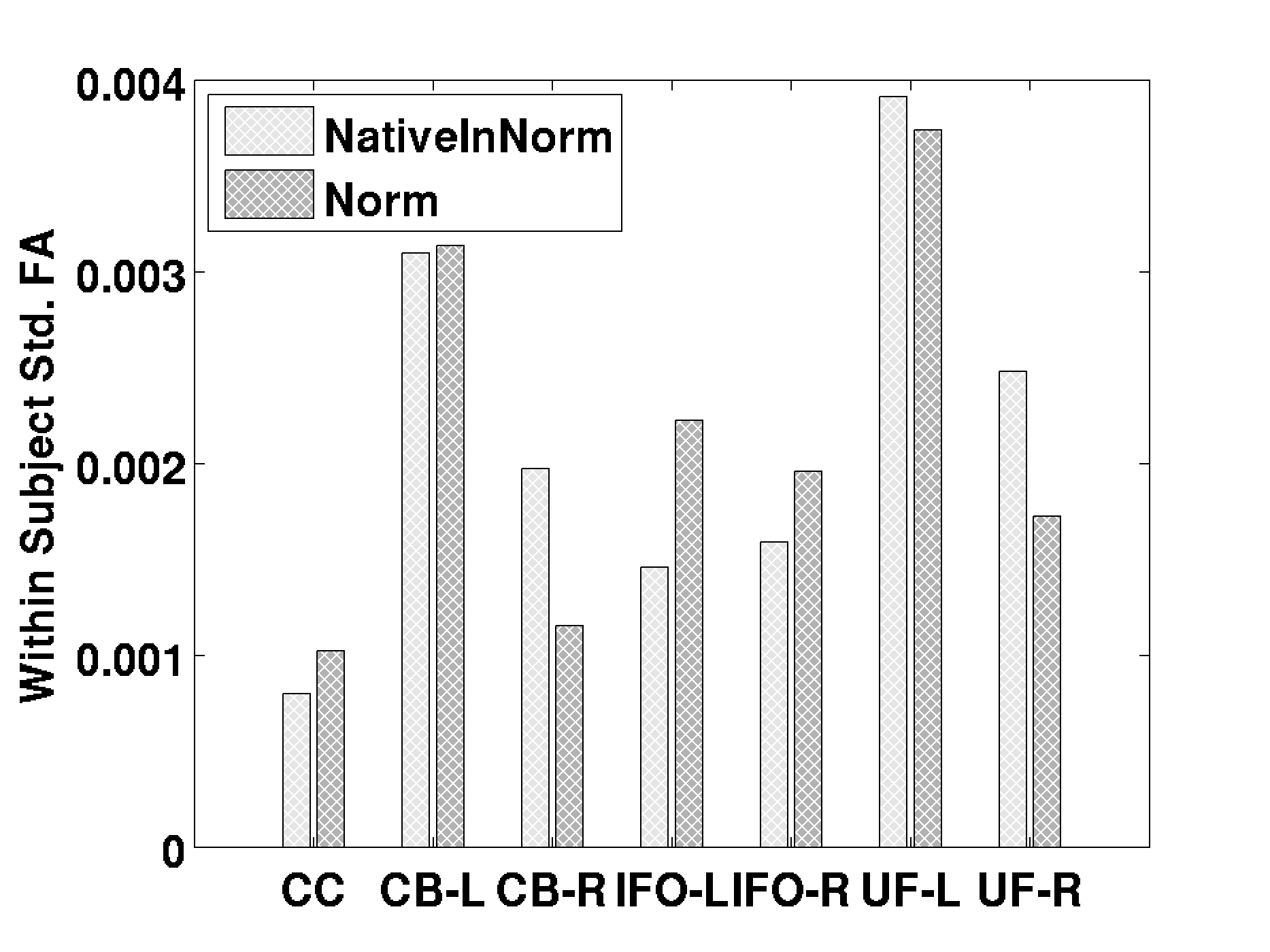}
\epsfxsize=0.33\linewidth\epsfbox{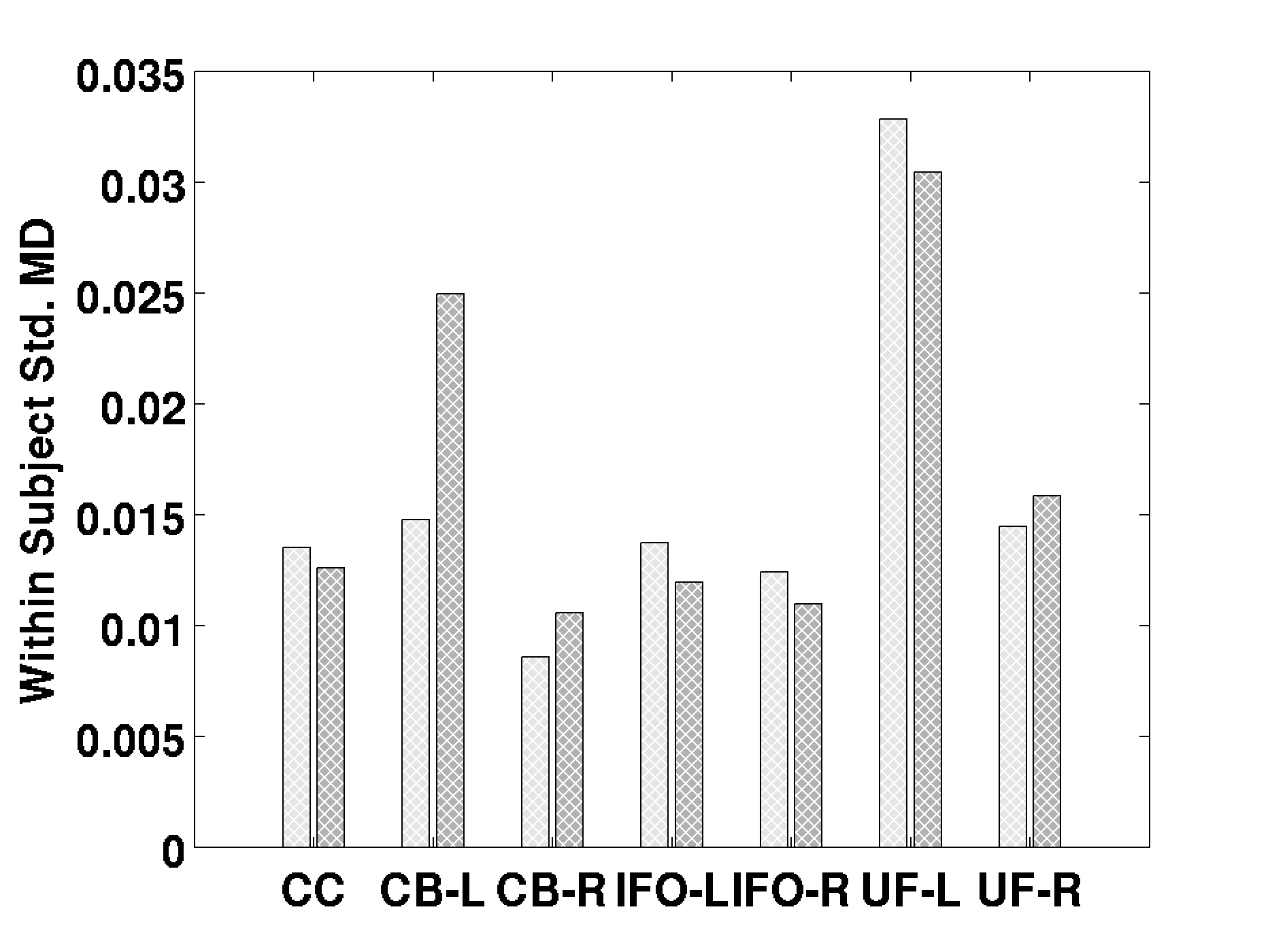}
\epsfxsize=0.33\linewidth\epsfbox{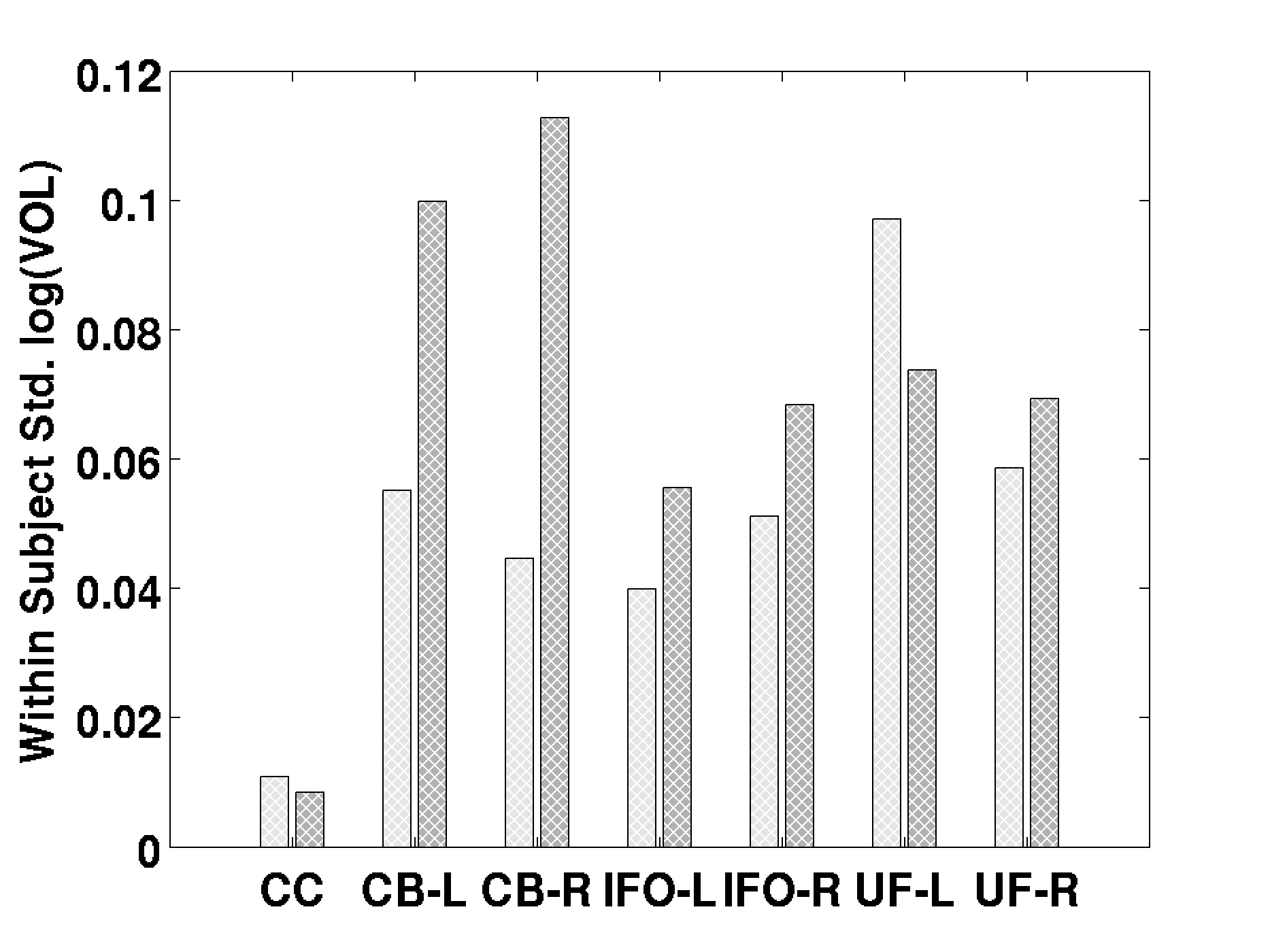}}
\caption{Within subject inter-scan standard deviations (computed using the individual repeat data) of the quantitative summaries of the tract reconstructions ($\overline{\mu}(\mathcal{T})$) of the four structures (including left and right side for CB, IFO and UF) obtained by tracking in non-linearly warped and native tensors. The units for MD and $\log$(VOL) are in $\upmu$m$^2$/s and mm respectively.}\label{fig:fa_tr_vol_within_subj_std}
\end{figure}

Fig. \ref{fig:fa_tr_vol_within_subj_std} presents the within subject standard deviations in the quantitative DTI and volumetric measures using the individual repeat data. We observe that the overall standard deviations are quite small and the tract reconstructions are consistent when using either unwarped or non-linearly warped tensors for tracking.

\begin{figure}[!htb]
\centerline{\epsfxsize=1.0\linewidth\epsfbox{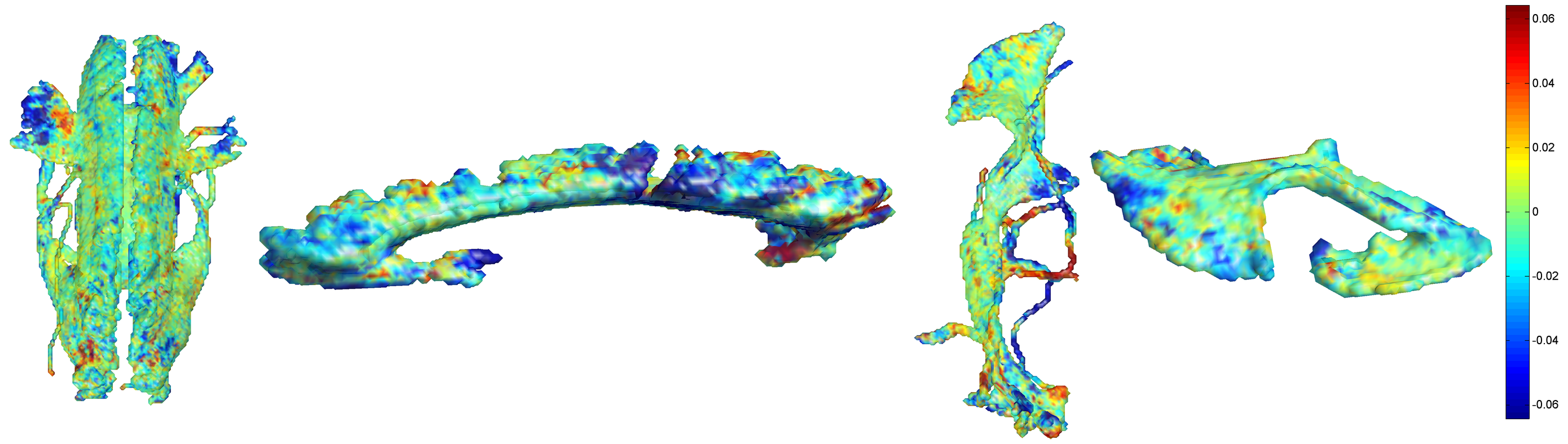}}
\centerline{\epsfxsize=1.0\linewidth\epsfbox{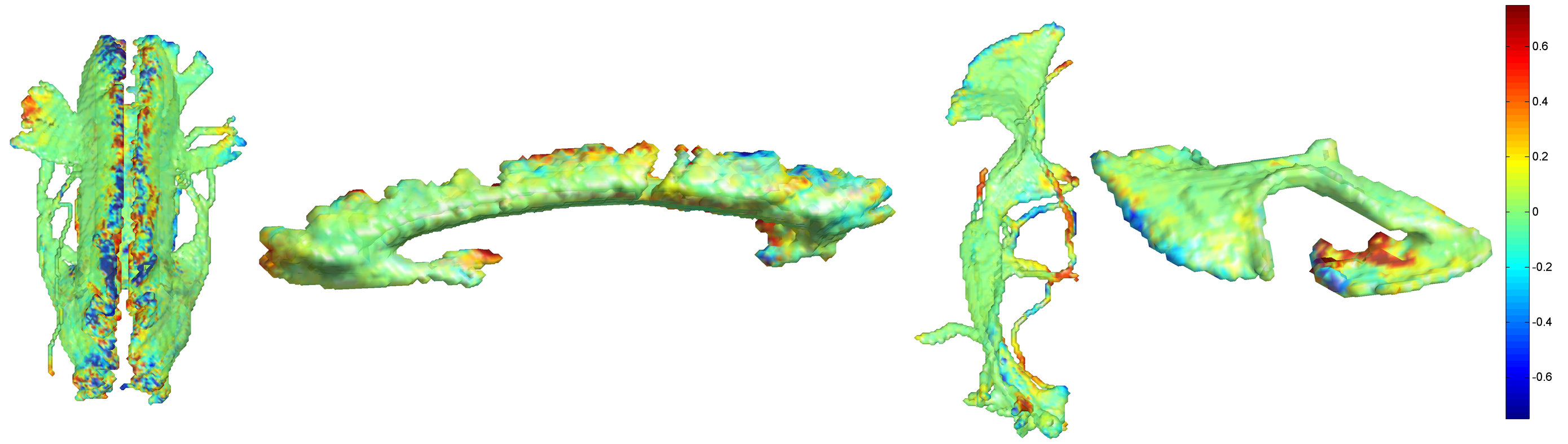}}
\caption{Microstructural consistency fluctuation maps. FA consistency fluctuations are shown on top and those of MD on the bottom. Left to right: CC suerior view, CB-left, IFO-left, UF-left. As we can see from the figures the fluctuations are minimal when considering the 0.06 range for FA and $\sim$ 0.6 $\upmu$m$^2$/s for MD.}\label{fig:fa_tr_surf}
\end{figure}

Finally, Fig. \ref{fig:fa_tr_surf} shows the microstructural consistency fluctuation maps. These maps are obtained similar to the shape consistency fluctuation maps but with minor technical differences in the computation of the fluctuations. For each vertex ($v_i^{\mathcal{A}}$) on the atlas tract surface the distance $d_i$ now is computed as follows,
\begin{equation}
d_i=\mathcal{M}\left(\underset{j}{\textrm{argmin}}~\{d(v_i^{\mathcal{A}},v_j^{\textrm{Norm}})\}_{j=1}^N\right)-\mathcal{M}\left(\underset{j}{\textrm{argmin}}~\{d(v_i^{\mathcal{A}},v_j^{\textrm{Native}})\}_{j=1}^{N'}\right).\label{eq:mcfm}
\end{equation}
The key differences between Eq. \eqref{eq:mcfm} and Eq. \eqref{scfm} are (1) the distances in the microstructural consistency fluctuation maps are signed to reflect whether normalized space tracking over or under estimates the microstructural properties compared to the native space and (2) first the vertex index is obtained via argmin operation and then the microstructural property is read off the FA or MD volume represented by $\mathcal{M}$. These maps show the differences in the measures FA and MD along the surface of the tracts mapped using native space and normalized space tractography. As we can observe the consistency fluctuates mainly along the borders. Please also see the supplementary material (\url{http://brainimaging.waisman.wisc.edu/~adluru/EvalTractography}) that has accompanying video that shows the fluctuations for different color scale mappings. These videos also show the inferior view of the CC and the CB, IFO and UF on the right hemisphere.

\section{Discussion}\label{sec:conclusion}
In this study we evaluated the effect of highly non-linear diffeomorphic warping of DTI on deterministic white matter fiber tracking. The data presented in this article indicate that the diffeomorphic warping with PPD based tensor reorientation \citep{zhang.2006} yields similar tractography reconstructions in both native and non-linearly warped DTI data. Large structures like $\mathtt{CC}$ are more consistent when using warped and unwarped tensors compared to long and narrow structures like $\mathtt{CB}$ and $\mathtt{IFO}$. In such cases there are some spurious branches when tracking in the warped tensors possibly due to smoothing effect and the worsened partial voluming of the $\mathtt{ROI}$s used to filter the pathways. For shorter tracts that are relatively isolated, such as the uncinate fasciculus the consistency is higher.

We would like to note that besides reorientation, another key component involved in DTI non-linear warping is tensor interpolation. It has been shown that the choice of metric for tensor interpolation (between Euclidean and log-Euclidean) makes a difference in the DTI metrics \citep{pasternak_et_al_2010}. However since log-Euclidean \citep{arsigny_et_al_2006} is shown to preserve the anisotropic components of the tensor which are important for tractography \citep{fillard_et_al_2007} we use log-Euclidean interpolation in our spatial warping. Studying the effects of different interpolations on tractography is an interesting question but is not considered in this study. Finally our results show the effect of warping with a \emph{composite} of rigid, affine and diffeomorphic transformations. As discussed in \citet{alexander_et_al_2001} tensor reorientation is perfect when using rigid-only transformations and diffeomorphic transformations are approximated using piecewise affine transformations. Hence we do not consider reporting individualized effects of rigid-only, affine-only transformations on tractography. This study also did not investigate the sensitivity to tract selection filter $\mathtt{ROI}$s or different types of tractography algorithms such as global \citep{kreher_et_al_2008} or probabilistic \citep{parker_et_al_pico_2003}. Although the probabilistic tractography can provide more data, the main conclusions of our results would be expected to extend for that framework of tracking as long as the algorithms are essentially “bootstrap” versions of deterministic and local streamline tractography. The global probabilistic tractography algorithms on the other hand form a different class of algorithms and is out of scope of our current investigation. We would also like to point out that such global methods are not widely used because of the computational burden.

We would also like to mention that there are various other factors that can influence tracking such as the stopping criteria used by thresholding FA, EPI distortion correction methods. But for the tracts considered in this paper we can observe that the tract shapes do not differ overall which means that the tract reconstructions are not effected by the stopping criteria. For more specific tracts such as centrum semiovale with a lot of fiber crossings, some special settings of stopping criteria might be more appropriate. But such criteria can be part of the "tract extraction protocol" and will not be a major factor in understanding the effects of \emph{spatial transformations}. Similarly, as presented in \citet{irfanoglu_et_al_2012}, fieldmap homogeneity and EPI distortions can have an impact on tractography however for the purpose of investigating the specific effects of spatial transformations this is not a major factor since the comparison is between pre and post spatial transformation \emph{after} the tensors are estimated. We would like to note again that as described in our paper, we did perform the standard field map correction using the FSL tool. This helps to understand the effects of spatial transformations in a standard practice setting.

Finally we acknowledge the limitations of the DT model but it is still one of the most widely used when it comes to diffusion imaging. Conceptually the accuracy of reorienting the higher-angular resolution models (e.g. using methods like \citep{hong_et_al_2009,raffelt_tournier_mrm_2011}) would affect the accuracy of tracking in the warped models. We would expect that the parts of the tracts that pass through crossing-fiber regions would be more effected. However exploring such models in detail is out of the scope of our current investigation.

\subsection{Implications for population studies}
Although voxel-based analyses (VBA) of DTI measures are attractive, they are typically limited in power because of the need to address multiple comparisons problem in controlling type I error or false positive rate. Hence the frameworks such as tract-based spatial statistics (TBSS) \citep{smith.2006} have become popular in WM studies. TBSS is useful in reducing the number of comparisons; however, it does not necessarily correspond to specific tracts and the regional mapping of measures is based upon the local maximum FA, which may obscure focal effects. Approaches such as region-of-interest analyses, provide an attractive alternative to test hypotheses about \emph{specific} WM substrates. Hence when trying to study the WM substrates of various cognitive and psychopathological human behaviors, tract-specific measures allow conducting statistically more powerful and computationally more efficient analyses.

Tractography facilitates more accurate localization of specific white matter structures which could be identified by following protocols such as \citep{wakana_et_al_2004,catani_et_al_2008}. Following such protocols in each individual subject data would be cumbersome in a population study. Hence often the filtering regions namely, way-point and exclusion regions, needed to identify the WM pathways are traced on a population specific atlas and then inverse warped into the subject space \citep{pagani_et_al_2005}. It has been indicated in \citet{mori_oishi_faria_2009} tractography based reconstructions of white matter structures using DTI could only be validated using strong \emph{a priori} knowledge about the shapes of the structures. Given such a limitation, the identification of the filtering regions in the atlas can be more effective when tractography data are available in the atlas which can provide real-time feedback of filtering by using the \emph{a priori} shapes of the structures. Population specific atlases benefit from averaging across many subjects in the normalized space and hence any the minor reorientation errors at the individual voxel level are expected to cancel out. Hence tractography on the atlas is performed in applications which perform tract specific statistical analyses \citep{yushkevich.2008,goodlett_neuro2009,zhang_et_al_2010}. These applications do not inverse warp the filtering regions but the binarized tractography masks \citep{yushkevich.2008,zhang_et_al_2010} or fiber tracts themselves \citep{goodlett_neuro2009} identified on the atlas.

Although there have been applications, such as above, that perform tractography on the atlas there are only a few articles that perform tracking in the non-linearly warped tensors of the \emph{individual} subjects for statistical analyses in the normalized space. Many approaches such as tractography-guided spatial statistics (TGIS) \citep{jungsu_et_al_2009} and tract based morphometry (TBM) \citep{odonnell_et_al_2009} however still prefer tracking in the unwarped tensors in the native space. Our data show that non-linear warping of DTI preserves the shapes of the tracts, suggesting that methods such as \citet{jungsu_et_al_2009} could obtain the 3-D parametric meshes of the tracts by directly tracking in the warped DTI in normalized space and thus achieve computational efficiency. In \citet{odonnell_et_al_2009} for example, warping of the tracts and the ill-posed tract clustering step could be avoided, since tracking in warped space readily establishes correspondences across subjects, needed for population analyses. Atlas based prior information could be also be utilized effectively by tractography algorithms when tracking in warped DTI data in an atlas space. Generating voxel-wise connectivity maps using tractography in normalized space can also become a feasible and provide options interesting analytic strategies. Without performing tracking in normalized space such an application would be extremely hard to implement.

Finally, we would like to note that although tractography has been used to measure and influence the quality of DTI warping \citep{zhang.2006,ziyan_et_al_2007,zollei_et_al_2010,zvitia_et_al_2010} and tractography reproducibility studies such as \citet{heiervang_et_al_2006,besseling_et_al_2012} have been performed, the work presented in this study addressed a distinct question: does non-linear warping of DTI with non-exact tensor reorientation preserve even long-range anatomical connections allowing simple deterministic tractography algorithms reconstruct white matter structures consistently in the normalized space. Our results quantitatively demonstrate that tracking fibers in the warped DTI produces anatomically consistent structures compared to tracking in unwarped native DTI, thus providing a firm ground for using tractography data at the individual level (not just at the average level) in population studies. Both shape and quantitative microstructural properties of tractography results are preserved during spatial normalization. Our data suggest that except when investigating extremely small effects on the white matter, quantitative properties of white matter structures could be extracted by performing tractography in warped DTI in the normalized space for performing regional, shape or topological differences between individuals and groups.
\section*{Acknowledgements}
We thank Frances Haeberli and Chad M. Ennis for assisting in the data collection and Steven R. Kecskemeti for helpful discussions. This work was supported by the NIMH R01 MH080826 and R01 MH084795 and the NIH Mental Retardation/Developmental Disabilities Research Center (MRDDRC Waisman Center), NIMH 62015, the Autism Society of Southwestern Wisconsin, the NCCAM P01 AT004952-04 and the Waisman Core grant P30 HD003352-45.
\bibliographystyle{model2-names}
\bibliography{reference,connectome,tractclustering,revision}

\end{document}